\documentclass{ws-ijmpa}
\usepackage{epsfig}
\usepackage{graphicx}

\newcommand{\Od}{{\cal O}}

\newcommand{\gsim}{\mbox{\raisebox{-.9ex}{~$\stackrel{\mbox{$>$}}{\sim}$~}}}
\usepackage[super,compress]{cite}
\begin{document}

\markboth{J. A. R. Cembranos and A. L. Maroto}
%{Disformal Scalars as Dark Matter Candidates: Branon Phenomenology}
{Disformal Scalars as Dark Matter Candidates: Branon Phenomenology}
\catchline{}{}{}{}{}
%
%%%%%%%%%%%%%%%%%%%%%%%%%%%%%%%%%%%%%%%%%%%%%%%%%%%%%%%%%%%%%%%%%%%%

%Phenomenology of Disformal Scalar Particles:
%Branon Dark Matter
\title{DISFORMAL SCALARS AS DARK MATTER CANDIDATES:
BRANON PHENOMENOLOGY}
%\title{BRANONS AS DARK MATTER CANDIDATES}
%\footnote{For the title, try not to use more than
%3 lines. Typeset the title in 10 pt roman, uppercase and
%boldface.}}

\author{JOSE A. R. CEMBRANOS}
%\footnote{Typeset names in 8 pt roman, uppercase. Use the footnote to indicate the
%present or permanent address of the author.}}
\address{
Departamento de  F\'{\i}sica Te\'orica I, Universidad Complutense de Madrid\\
Madrid, Spain E-28040,
\\
cembra@fis.ucm.es}

\author{ANTONIO L. MAROTO}
\address{
Departamento de  F\'{\i}sica Te\'orica I, Universidad Complutense de Madrid\\
Madrid, Spain E-28040,
\\
maroto@fis.ucm.es}

\maketitle

\begin{history}
\received{Day Month Year}
\revised{Day Month Year}
\end{history}

\begin{abstract}
Scalar particles coupled to the Standard Model fields through a disformal coupling
arise in different theories, such as massive gravity or brane-world models.
We will review the main phenomenology associated with such particles.
Distinctive disformal signatures could be measured at colliders and with astrophysical observations.
The phenomenological relevance of the disformal coupling demands the introduction of
a set of symmetries, which may ensure the stability of these new degrees of freedom.
In such a case, they constitute natural dark matter candidates since they
are generally massive and weakly coupled. We will illustrate these ideas by paying particular attention
to the branon case, since these questions arise naturally in brane-world models with
low tension, where they were first discussed.
\end{abstract}

\ccode{PACS numbers:95.35.+d, 11.25.-w, 11.10.Kk}
%•	95.35.+d Dark matter
%•	11.25.-w Strings and branes
%•	11.10.Kk Field theories in dimensions other than four

\section{Introduction}
Disformal scalar fields were introduced by Bekenstein~\cite{Bekenstein:1992pj}. By assuming the weak equivalence principle and causality,
he showed that a scalar field $\pi$ can modify the geometry by defining the following metric $g_{\mu\nu}$:
\begin{equation}
g_{\mu\nu}=A(\pi,X)\tilde{g}_{\mu\nu} + B(\pi,X) \partial_\mu \pi \partial_\nu \pi\; ,
\label{eq:bekmetric}
\end{equation}
where $\tilde{g}_{\mu\nu}$ is a background metric independent of the scalar field, and
$X=(1/2)\tilde{g}^{\mu\nu}\partial_{\mu}\pi\partial_{\nu}\pi$ depends on the first derivatives of such a field.
This is the most general metric that respects the previous two basic constraints.
If $B(\pi,X)=0$, Eq. (\ref{eq:bekmetric}) defines a conformal relation between $g_{\mu\nu}$ and $\tilde{g}_{\mu\nu}$.
In such a case, the scalar field is named conformal, and the conformal factor $A(\pi,X)$ defines the conformal
coupling of $\pi$ to the rest of fields. Conformal scalar fields have been studied from a long time and its phenomenology
can be found in a large number of works from different approaches.

On the other hand, the term proportional to $B(\pi,X)$ is the genuine disformal coupling. The presence of a non-zero
disformal factor $B(\pi,X)$ is able to change dramatically the phenomenology of the scalar field. We will focus on such a case
by assuming that the disformal coupling dominates over the conformal one. The simplest possibility for
the disformal coupling is just a constant term $B(\pi,X)\propto 1/f^4$, where $f$ has dimensions of energy.  In such a case the main interaction of $\pi$
with the rest of fields is through their energy momentum tensor $T^{\mu\nu}$:
\begin{equation}
\mathcal{L}_D \propto \frac{1}{f^4}\partial_\mu \pi\partial_\nu \pi T^{\mu\nu}\;.
\label{eq:coupling}
\end{equation}
Eq. (\ref{eq:coupling}) shows explicitly the basic property of the disformal model. It is an effective field theory governed by an energy dimension 8 operator. The high power of the leading interaction may seem artificial and unstable against radiative corrections. Indeed, this is the case except if additional symmetries are present.
First of all, the derivative coupling shown in
(\ref{eq:coupling}) preserves the shift symmetry of the field: $\pi(x^\mu)\rightarrow\pi'(x^\mu)\equiv \pi(x^\mu)+\Lambda$,
where $\Lambda$ is a constant. In this sense, the disformal scalar can be associated with the Nambu-Goldstone Boson (NGB) arising from the spontaneous breaking of a global symmetry. If the symmetry is exact, the $A$ and $B$ functions can only depend on the $X$ term:
$A(\pi,X)=A(X)$ and  $B(\pi,X)=B(X)$. On the other hand, the symmetry can be slightly violated in an explicit way. In this case, the leading phenomenology can still be associated with the term described in Eq. (\ref{eq:coupling}), and the disformal scalar can be understood as a pseudo-Nambu-Godstone Boson (pNGB).

If the shift symmetry is not exact, interaction terms with a single scalar field are expected. If they are not present in the tree level Lagrangian, they will generally arise due to radiative corrections. However, Eq. (\ref{eq:coupling}) owns an additional discrete global symmetry. It is invariant under the parity transformation of the disformal field: $\pi(x^\mu)\rightarrow\pi'(x^\mu)\equiv -\pi(x^\mu)$. If this symmetry is imposed, terms with an odd number of disformal scalars can be forbidden. In this case, the scalars will be coupled in pairs to the SM particles. This symmetry implies a complete different phenomenology for disformal fields than for standard dilatons. In particular, this type of disformal scalars are stable and can play an important role in cosmology.

The first detailed framework where these ideas were developed was in flexible brane-world models. These models are characterized by the fact that Standard Model (SM) particles are restricted to propagate on a manifold of three spatial dimensions embedded in a higher dimensional space-time ($D=4+N$). Only the gravitational interaction has access to the whole bulk space.
The fundamental scale of gravity  in $D$ dimensions, $M_D$, can be reduced in relation to Planck scale $M_P$, due to a large volume of the extra-space. In the original proposals, the value of $M_D$ was taken around the electroweak scale since they try to address the hierarchy problem
\cite{ADD}. However, the model suffers from important constraints, and $M_D$ has to be much larger \cite{Langlois,six}.
From an observer on the brane, the existence of extra dimensions introduces new degrees of freedoms, that can be studied within an
effective field theory at low energies. On the one hand, modes of fields propagating into the bulk space have associated a so-called Kaluza-Klein (KK) tower of states. In principle, this tower will be restricted to gravitons, but more complex models can have  other degrees of freedom with access to the bulk space. On the other hand, the fluctuations of the brane will be parameterized by several $\pi^\alpha$ fields of spin zero.
The flexible character of the brane is quantified by the brane tension $\tau\equiv f^4$. These scalars are
called branons. If the translational invariance along the extra-dimensions is an exact symmetry, they can be understood as the massless
NGB arising from the spontaneous breaking of that symmetry induced by the location of the brane in a particular point of the extra-space \cite{Sundrum,DoMa}. In a more general case, a non-trivial energy content along the bulk space will explicitly break the translational invariance in the extra-space. In such a case, branons are expected to be massive \cite{BSky,Alcaraz:2002iu}.

It is interesting to note that flexible branes suppress exponentially the coupling of the SM particles to any KK mode \cite{GB}. Therefore, if the tension scale $f$ is much smaller than the fundamental scale of gravity $M_D$, the KK states decouple from the SM particles. In such a case, the constraints on the model \cite{KK} are strongly alleviated. In fact, for flexible enough branes, the only relevant new particles at
low energies are the branon fields.

Branons have the standard disformal coupling given by Eq. (\ref{eq:coupling}). The potential signatures in colliders have been studied in different works. The general massive case was first discussed in Ref. \citen{Alcaraz:2002iu}, whereas the massless one was studied previously in
Ref. \citen{strumia}.
The force mediated by disformal scalars can be found in Refs. \citen{thesis} and \citen{Kugo} for the massive and massless scalars.
Limits from supernovae and modifications of Newton's law at small distances in the massless case were obtained in Ref. \citen{Kugo}. Moreover, in Refs. \citen{CDM} and \citen{CDM2}, the interesting possibility that massive disformal branons could account for the observed Dark Matter (DM) of the universe was studied in detail. From a more general disformal model approach, the constraints have been also studied in different contexts \cite{Kaloper:2003yf}.
In particular, disformal interactions arise in galileon models and massive gravitational theories \cite{deRham:2010ik,deRham:2010kj}.
 Within these frameworks, the work has focused on astrophysical and cosmological analyses \cite{Zumalacarregui:2010wj,Koivisto:2012za,Bettoni:2012xv,Brax:2013nsa,vandeBruck:2013yxa,Neveu:2014vua}, but laboratory \cite{Brax:2012ie} and fundamentally collider experiments \cite{Brax:2014vva} are presently the most constraining.

This manuscript is organized as follows: in
Section \ref{BraMo} we give a brief introduction to the branon model as an explicit example
of massive disformal fields.
Section \ref{Coll} contains a summary of the main disformal signatures at colliders.
Section \ref{Rel} is devoted to the standard calculation of relic abundance of
disformal scalars generated by the freeze-out phenomenon in an
expanding universe.
The constraints derived from this computation are presented in Section \ref{RelCon}.
In Section \ref{Dir} and \ref{Ind}, we analyze the possibility of detecting disformal DM
through direct and indirect searches respectively. On the other hand,
light disformal scalars suffer other types of astrophysical constraints.
In Section \ref{BBNs}, after  reviewing  the limits imposed by nucleosynthesis
on the number of relativistic species, we apply them to the disformal case.
Section \ref{Ste} contains an estimation for the rate of energy loss from stellar
objects in the form of disformal scalars and the corresponding constraints. In Section \ref{nonthermals} the possibility
for disformal DM to be produced non-thermally is explored.
Section \ref{Con} includes the main conclusions of our discussion. This
manuscript is completed with the addition of three appendices. \ref{Ver}
contains the disformal vertices with SM particles. \ref{Cross} includes the
formulas for the creation and annihilation cross-sections for disformal
scalar pairs. Finally, we give the results for the thermal averages of
disformal particle annihilation cross-sections into SM particles in \ref{Thermal}.

\section{Branon model}
\label{BraMo}
In this Section, we will describe in detail a particular model of disformal scalar
fields. We will introduce the main properties of massive brane fluctuations
in brane-world models \cite{DoMa,BSky,Alcaraz:2002iu}.
The original concept is associated with a single brane in the thin limit,
although the branon field can be more generally understood as a particular
coherent mode. The standard four-dimensional space-time $M_4$ is assumed to
be embedded in a bulk space of $D$ dimensions. For simplicity, we will assume it to
be of the form $M_D=M_4\times B$, i.e. we can define the extra space $B$,
as an independent $N$-dimensional compact manifold, so $D=4+N$. The brane will lie along
the $M_4$ space-time, and we will work in the probe brane limit
neglecting its contribution to the bulk gravitational field.
The coordinates $(x^{\mu},y^m)$ parameterize the points in the bulk space $M_D$,
where the indices run as $\mu=0,1,2,3$ and $m=1,2,...,N$. $M_D$ is endowed with a
metric $G_{MN}$ with signature $(+,-,-...-,-)$. For simplicity, we will
assume the separable geometry defined by the following ansatz:
\begin{eqnarray}
 G_{MN}&=&
\left(
\begin{array}{cccc}
\tilde g_{\mu\nu}(x,y)&0\\ 0&-\tilde g'_{mn}(y)
\end{array}\right).
\end{eqnarray}
The position of the brane in the bulk space-time $M_D$ can be parameterized naturally with the
coordinates of $M_4$: $Y^M=(x^\mu, Y^m(x))$, with $M=0,\dots, 3+N$, so the first four
coordinates have been chosen to be identified with the space-time brane coordinates $x^\mu$.
In this way, the brane is located in a particular point in the extra space $B$,
i.e. $Y^m(x)=Y^m_0$. This position defines its ground state. We will consider
that the extra space is homogeneous, so that brane fluctuations can be written in terms
of properly normalized coordinates in $B$: $\pi^\alpha(x)=f^2 Y^\alpha(x)$, where
$\alpha=1,\dots, N$. The geometry that determines the dynamics on the brane is defined
by the induced metric. In the ground state, this metric is simply given by the four-dimensional
components of the bulk space-time metric: $g_{\mu\nu}=\tilde g_{\mu\nu}=G_{\mu\nu}$. However, in general,
brane fluctuations will modify it in the following way
\begin{eqnarray}
g_{\mu\nu}=\partial_\mu Y^M\partial_\nu Y^N G_{MN}(x,Y(x)) =\tilde
g_{\mu\nu}(x,Y(x))-\partial_{\mu}Y^m\partial_{\nu}Y^n\tilde
g'_{mn}(Y(x))\,.
\label{induced}
\end{eqnarray}
The induced metric (\ref{induced}) can be expanded around the ground state
in order to find explicitly the branon contributions \cite{DoMa,BSky,Alcaraz:2002iu} :
\begin{equation}
g_{\mu\nu}=
\tilde g_{\mu\nu}-\frac{1}{f^4}\delta_{\alpha\beta}\partial_{\mu}\pi^\alpha
\partial_{\nu}\pi^\beta
+\frac{1}{4f^4}\tilde g_{\mu\nu}M_{\alpha\beta}^2\pi^\alpha\pi^\beta
+\dots
\end{equation}
Branons can be defined as the mass eigenstates of the brane fluctuations within the extra-space directions.
The matrix $M_{\alpha\beta}$ determines the branon masses. It characterizes the local geometrical properties of the bulk space
where the brane is located. In the absence of an explicit model for the bulk dynamics, its elements should be
considered as free parameters (for instance, for a particular construction, see Ref. \citen{Andrianov}).
Branons have zero mass only in highly symmetric bulk spaces \cite{DoMa,BSky,Alcaraz:2002iu}.
There is also the possibility of having models with massless and massive branons.
It will mean that the extra space has an incomplete set of isometries,
since the isometries have associated zero eigenvalues of $M_{\alpha\beta}$, being
the massless branons, the fields that parameterize the corresponding flat directions.

In this review, we will not discuss the fundamental nature of the brane. On the contrary, we will assume
that its dynamics can be described by a low-energy effective action \cite{DoMa,BSky,Alcaraz:2002iu}.
In particular, we will consider that the kinetic term comes from the Nambu-Goto action and will take the
limit in which gravity decouples $M_D\rightarrow \infty$, since in such a case,
branon effects can be analyzed independently.

On the other hand, branon couplings to the SM particles
can be obtained from the standard action on a curved background
geometry given by the induced metric (\ref{induced}), which can be expanded
in branon fields. For example, the complete action up to second
order contains the SM terms, the kinetic term for branons and the
quadratic interaction term between branons and SM fields:
\begin{eqnarray}
S_B&=& \int_{M_4}d^4x\sqrt{g}[-f^4+ {\mathcal L}_{SM}(g_{\mu\nu})]
\nonumber\\
&=&\int_{M_4}d^4x\sqrt{\tilde g}\left[-f^4+ {\mathcal L}_{SM}
( \tilde g_{\mu\nu})  +
\frac{1}{2}\tilde g^{\mu\nu}\delta_{\alpha\beta}\partial_{\mu}
\pi^\alpha
\partial_{\nu}\pi^\beta-\frac{1}{2}M^2_{\alpha\beta}
\pi^\alpha\pi^\beta\right.
\nonumber\\
&+&
\left.\frac{1}{8f^4}(4\delta_{\alpha\beta}\partial_{\mu}\pi^\alpha
\partial_{\nu}\pi^\beta-M^2_{\alpha\beta}\pi^\alpha\pi^\beta
\tilde g_{\mu\nu})
T^{\mu\nu}_{SM}(\tilde g_{\mu\nu}) \right]
+\dots \label{lag}
\end{eqnarray}
Here, $T^{\mu\nu}_{SM}(\tilde g_{\mu\nu})$ is the conserved
energy-momentum tensor of the SM evaluated in the
background metric $\tilde g_{\mu\nu}$:
\begin{eqnarray}
T^{\mu\nu}_{SM}=-\left(\tilde g^{\mu\nu}{\mathcal L}_{SM}
+2\frac{\delta {\mathcal L}_{SM}}{\delta \tilde
g_{\mu\nu}}\right)\,.
\end{eqnarray}

The quadratic expression in (\ref{lag}) is general for any extra space $B$,
regardless of the form of the metric $\tilde g'_{mn}$.
Indeed the low-energy effective Lagrangian is model independent and is parameterized only by
the number of branon fields, their masses and the brane tension.
The particular geometry of the total bulk space only affects at higher orders.
Therefore, these effective couplings provide the necessary tools to analyze the phenomenology
of branons in terms of $f$ and the branons masses.

It is interesting to note that under a parity transformation in the bulk space,
the extra dimensions change sign. As they are parameterized by a the branon fields,
these fields changes sign as well. The symmetry under extra dimensional parity
is preserved by the above Lagrangian (\ref{lag}).
However, in a general case, this parity may be violated by higher orders
terms. As we have commented in the introduction, this violation may introduce
interacting terms in the Lagrangian with a single branon field. For this reason,
we will restrict our analysis to bulk geometries that are invariant under
parity transformations in the extra dimensions around the brane location.
This is the case of the most part of the brane-world models. In particular,
it is well known that in order to introduce quiral fermions on the bulk space,
the common $S^1$ symmetry should be promoted to a $S^1\times \mathbb{Z}_2$,
i.e. it has to be defined in an orbifold. This new symmetry, that we will call
brane-parity ensures the absence of terms in the effective Lagrangian with an odd number of
branons. In such a case, branons interact by pairs with the SM particles and are necessarily stable.
On the other hand, branon couplings are suppressed by the brane tension scale
$f$, which means that they may be weakly interacting. As we have discussed above,
they are generally massive. Therefore, their freeze-out temperature can be relatively high and their
relic abundances can be cosmologically important. We will discuss the phenomenology of
branons in laboratories and cosmology as a paradigmatic example of disformal scalar fields.

\section{Disformal fields at colliders}
\label{Coll}

Independently of the cosmological or astrophysical importance of disformal scalar fields,
they can be searched for in collider experiments. In fact, the parameters of the Lagrangian (\ref{lag})
suffer the constraints from present observations. On the other hand, they may be detected at the LHC
or in a future generation of colliders
\cite{Alcaraz:2002iu,Brax:2014vva,Coll,L3,Cembranos:2004jp,LHCDirect,Landsberg:2015pka,Khachatryan:2014rwa,BWRad}.

We will pay attention to the most sensitive searches that can be performed
at the LHC, but we will summarize a large amount of studies at the end
of this section and in Tables \ref{tabHad} and \ref{radcoll}.
The most sensitive production process for disformal particles in a proton-proton
collider, as the LHC, is gluon fusion giving a gluon in addition to
a disformal particle pair; and the quark-gluon interaction giving a
quark and the mentioned pair. These processes contribute to the
monojet $J$ plus transverse missing energy and momentum signature.

An additional and complementary process is the quark-antiquark annihilation giving a photon and a
pair of disformal particles.
In such a case, the signature is a single photon in addition to the transverse missing energy and momentum.
In order to show the constraints, we will restrict the study to a model with a degenerated spectrum of $N$ disformal fields with a common mass $M$. For simplicity, we will consider massless quarks except for the top case.
The cross-section of the subprocess $g g \rightarrow g \pi\pi$ was computed in
Refs. \citen{Cembranos:2004jp} and \citen{LHCDirect}:
\begin{eqnarray}
&&\frac{d\sigma(g g \rightarrow g\pi\pi)}{dk^2dt}=\frac{\alpha_s N
(k^2-4M^2)^2}{40960f^8\pi^2\hat s^3tu}\sqrt{1-\frac{4M^2}{k^2}} \nonumber  \\
&&(\hat s^4+t^4+u^4-k^8+6k^4(\hat s^2+t^2+u^2)-4k^2(\hat
s^3+t^3+u^3))\,.
\end{eqnarray}
Here, $\hat s\equiv(p_1+p_2)^2$, $t \equiv(p_1-q)^2$, $k^2\equiv(k_1+k_2)^2$ and
 $u\equiv(p_2-q)^2$. $p_1$ and $p_2$ are the initial gluon four-momenta, whereas
 $q$ is the final gluon four-momentum. Finally, $k=k_1+k_2$ is the total branon four-momentum.
Thus, the contribution from this subprocess to the total cross-section for the process $p
p\rightarrow g\pi\pi$ can be written as
\begin{eqnarray}
\sigma_{gg}(p p\rightarrow g\pi\pi)= \int_{x_{min}}^1
dx\int_{y_{min}}^1 dy
g(y;\hat s) g(x;\hat s)   \nonumber\\
\int_{k^2_{min}}^{k^2_{max}}  dk^2 \int_{t_{min}}^{t_{max}}
dt\frac{d\sigma(g g\rightarrow g\pi\pi)}{dk^2dt},
\end{eqnarray}
where $g(x;s)$ is the gluon distribution function associated to the proton, $x$
and $y$ are the energy fractions of the protons carried by the two
initial gluons. The integration limits depend on the cuts used to define
 the total cross-section. For example, the identification of a monojet forces to
 consider a minimal value for its transverse energy $E_T$ and a pseudorapidity range:
 $(\eta_{min},\eta_{max})$. It implies the limits
 $k^2_{min}=4M^2$, $k^2_{max}=\hat s(1-2E_T/\sqrt{\hat s})$ and
 $t_{min(max)}=-(\hat s-k^2)[1+\tanh{(\eta_{min(max)})}]/2$ . In addition, we can define
 $x_{min}=s_{min}/s$ and $y_{min}=x_{min}/x$ where $s$ is the total center of mass energy
 squared of the process. It has also a minimum value:
\begin{equation}
s_{min}=2E_T^2+4M^2+2E_T\sqrt{E_T^2+4M^2}.
\end{equation}
Similarly, the $q g \rightarrow q \pi\pi $ subprocess is characterized by the following cross-section:
\cite{Cembranos:2004jp,LHCDirect}
\begin{eqnarray}
&&\frac{d\sigma (qg\rightarrow q\pi \pi )}{dk^{2}dt}  \nonumber \\
&=&-\frac{\alpha _{s}N}{2}\frac{(k^{2}-4M^{2})^{2}}{184320f^{8}\pi ^{2}\hat{s%
}^{3}tu}\sqrt{1-\frac{4M^{2}}{k^{2}}}(uk^{2}+4t\hat{s})(2uk^{2}+t^{2}+\hat{s}%
^{2})\,.
\end{eqnarray}%
In this case, $p_1$ and $p_2$ are the quark and the gluon four-momenta
respectively. $q$ is the four-momentum associated to the final quark.
$k_1$ and $k_2$ are again the branon four-momenta. The Mandelstam variables can
be defined as in the previous case.

The total cross-section for the reaction $p p\rightarrow q\pi\pi$ can be written as
\begin{eqnarray}
\sigma(p p\rightarrow q\pi\pi)= \int_{x_{min}}^1 dx\int_{y_{min}}^1
dy \sum_q
 g(y;\hat s) q_p(x;\hat s)   \nonumber\\
\int_{k^2_{min}}^{k^2_{max}}  dk^2 \int_{t_{min}}^{t_{max}}
dt\frac{d\sigma(q g\rightarrow q\pi\pi)}{dk^2dt}\,.
\end{eqnarray}
Here, $x$ and $y$ are the energy fractions of the two protons
carried by the quark and the gluon associated to the subprocess,
and the limits of the integrals can be deduced as in the previous case
\cite{Cembranos:2004jp,LHCDirect}.

The above equations can be used to compute the total cross-section for
monojet production depending on the cut in the jet transverse energy $E_T$.

As we have commented, an interesting complementary signature is provided by the
single photon channel. The cross-section of the subprocess $q \bar q
\rightarrow \gamma \pi\pi$ was also computed in Refs.  \citen{Cembranos:2004jp}
and \citen{LHCDirect}:
\begin{eqnarray}
&&\frac{d\sigma (qq\rightarrow \gamma \pi \pi )}{dk^{2}dt}  \notag \\
&=&\frac{Q_{q}^{2}\alpha N(k^{2}-4M^{2})^{2}}{184320f^{8}\pi ^{2}\hat{s}%
^{3}tu}\sqrt{1-\frac{4M^{2}}{k^{2}}}(\hat{s}k^{2}+4tu)(2\hat{s}%
k^{2}+t^{2}+u^{2})\,.
\end{eqnarray}%
This analysis is simpler since it determines the only leading contribution:
\begin{eqnarray}
\sigma(p p\rightarrow \gamma\pi\pi)= \int_{x_{min}}^1
dx\int_{y_{min}}^1 dy \sum_q
 \bar q_{ p}(y;\hat s) q_{ p}(x;\hat s)   \nonumber\\
\int_{k^2_{min}}^{k^2_{max}}  dk^2 \int_{t_{min}}^{t_{max}}
dt\frac{d\sigma(q q \rightarrow \gamma\pi\pi)}{dk^2dt}\,.
\end{eqnarray}
The above cross-sections allow to compute the expected number
of events produced at the LHC for the mentioned channels. This number
depends on the disformal coupling $f$, the disformal mass $M$, and the number of
disformal scalars $N$.

%%%%%%%%%%%%%%%%%%%%%%%%%%%%%%%%%%%%%%%%%%%
%% Table 1
%%%%%%%%%%%%%%%%%%%%%%%%%%%%%%%%%%%%%%%%%%%
\begin{table}%[bt]
\tbl{
Summary of direct searches of disformal particles at colliders (results at the $95\;\%$ c.l.).
Monojet and single photon analyses are labeled by the upper indices ${1,2}$, respectively.
Present constraints and prospects for the LHC \cite{Landsberg:2015pka,Khachatryan:2014rwa,Cembranos:2004jp,LHCDirect}
are compared with current limits from LEP \cite{Alcaraz:2002iu,L3}, HERA and Tevatron \cite{Cembranos:2004jp}.
$\sqrt{s}$ means the center of mass energy associated with the total process;
${\mathcal L}$ denotes the total integrated luminosity;
 $f_0$ is the limit on the coupling for one massless disformal particle ($N=1$)
 and $M_0$ is the constraint on the disformal mass in the decoupling limit $f\rightarrow0$.
The effective disformal action (\ref{lag}) is not valid for energy scales
$\Lambda\gtrsim 4\,\pi^{1/2}f\,N^{-1/4}$ \cite{BWRad}. Thus, the $M_0$ is a limit that
cannot be reached but characterizes the sensitivity of the analysis for heavy disformal particles.
}
{%\small{
\begin{tabular}{||c|cccc||}
%\hline
\hline Experiment
&
% Process &
$\sqrt{s}$(TeV)& ${\mathcal
L}$(pb$^{-1}$)&$f_0$(GeV)&$M_0$(GeV)\\
\hline
%
%\vspace*{-.0cm} &&&\\
%\vspace*{.1cm}
%
HERA$^{\,1}
%{UCH}
$& 0.3 & 110 &  16 & 152
\\
Tevatron-II$^{\,1}
%{UCH}
$& 2.0 & $10^3$ &  256 & 902
\\
Tevatron-II$^{\,2}
%\gamma
$& 2.0 & $10^3$ &   240 & 952
\\
LEP-II$^{\,2}
%{UCH}
$& 0.2 & 600 &  180 & 103
\\
%
%\hline
%
%
LHC$^{\,2}
%\gamma
$& 8 & $19.6\times10^{3}$ &   440 & 3880
\\
\hline
LHC$^{\,1}
%{UCH}
$& 14 & $10^5$ &  1075 & 6481
\\
LHC$^{\,2}
%\gamma
$& 14 & $10^5$ &   797 & 6781
\\
\hline
\end{tabular}
}\label{tabHad}
\end{table}

\begin{figure}[bt]
\begin{center}
%\resizebox{8.8cm}{6.4cm}
\resizebox{8.5cm}{!}
{\includegraphics{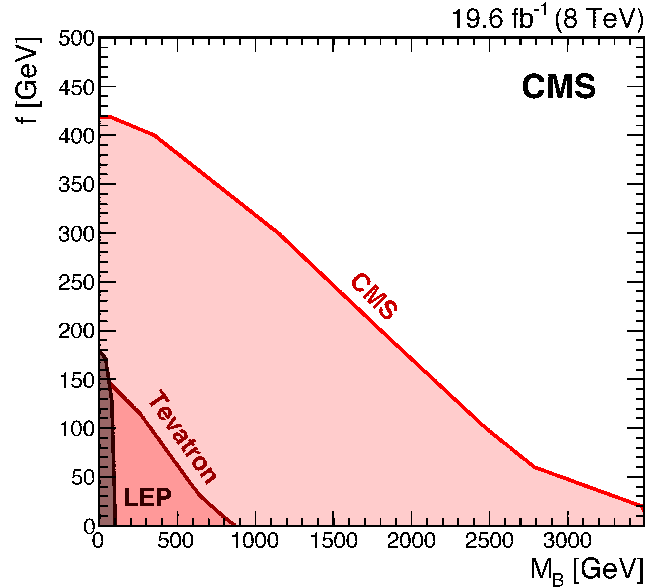}}
\caption {\footnotesize Present constraints for one disformal scalar particle obtained by the CMS collaboration.
This figure is taken from a dedicated analysis for  branon fields \cite{Khachatryan:2014rwa,Landsberg:2015pka}.
The results are compared with the restrictions coming from Tevatron \cite{Cembranos:2004jp} and LEP \cite{L3}.\label{CMSbranons}}
\end{center}
\end{figure}

In addition to these channels, there are other constraints on the
disformal model parameters for tree-level
processes from analyses of other collider signatures. A summary of
these studies can be found in Table \ref{tabHad} \cite{BW2,Coll,Cembranos:2004jp,LHCDirect}.
In this Table, the present restrictions coming from HERA, LEP-II and Tevatron
are compared with the present bounds of the LHC running at a center of mass energy
(c.m.e.) of 8 TeV and the prospects for the LHC running at 14 TeV c.m.e.
with full luminosity. For the single photon channel, CMS has reported a
dedicated analysis for branons with a total integrated luminosity of 19.6 fb$^{-1}$,
whose results can be observed in Fig. \ref{CMSbranons}.
\cite{Khachatryan:2014rwa,Landsberg:2015pka}. Other missing energy and transverse momentum
processes, such as those associated with the monolepton channel analyzed in Ref. \citen{Brax:2014vva}, are
also potential signatures of disformal models. In the same reference, the authors
develop an interesting discussion of other different phenomenological signals of these models.
However, the important dependence of the interaction with the energy makes them subdominant with respect to the
collider constraints.

On the other hand, it has been shown that disformal radiative corrections
generate new couplings among SM particles, which can be described by an
effective Lagrangian. The most relevant terms of these
effective interactions can be written as \cite{BWRad}:
\begin{eqnarray}\label{eff}
{\mathcal L}_{SM}^{(1)}\simeq \frac{N \Lambda^4}{192(4\pi)^2f^8}
\left\{2T_{\mu\nu}T^{\mu\nu}+T_\mu^\mu T_\nu^\nu\right\}\,.
\end{eqnarray}
The $\Lambda$ parameter is introduced when dealing with disformal radiative
corrections since the Lagrangian (\ref{lag}) is not renormalizable.
This parameter can be understood as a cutoff that limits
the energy range where the effective description of the disformal field
and SM particles is valid. In this case, from our approach, $\Lambda$ is
a phenomenological parameter, and $\Lambda/f$ parameterizes the intensity
of the quantum effects within the disformal theory. So, it quantifies the
relative importance of tree-level versus loop effects, which in principle, is unknown.
As it was shown in Ref. \citen{BWRad},
a perturbative treatment of the loop expansion is only consistent if $\Lambda\,<\,4\sqrt{\pi}fN^{-1/4}$.

In addition, one-loop effects computed
within the new effective four-fermion vertices coming from
(\ref{eff}) can be shown to be equivalent to the two-loop phenomenology
associated with the original action given by (\ref{lag}). For example,
this analysis can be used to obtain the contribution of disformal scalars
to the anomalous magnetic moment of the muon \cite{BWRad}:
\begin{equation}\label{gb}
\delta a_\mu \simeq \frac{5\, m_\mu^2}{114\,(4\pi)^4}
  \frac{N\Lambda^6}{f^8}
\end{equation}
where $N$ is the number of disformal fields. On the other hand,
the most important disformal radiative effects in SM phenomenology
at colliders are related to four-fermion interactions or
different fermion anti-fermion pair annihilation into two gauge bosons
\cite{BWRad}. By considering current data, it is possible to constrain
the particular combination of parameters $f^2/(\Lambda N^{1/4})$.
Present results are shown in Table \ref{radcoll}, where it is also possible to find the
prospects for the LHC running at $14$ TeV \cite{BWRad} and current constraints obtained with data coming from HERA
\cite{Adloff:2003jm}, Tevatron \cite{d0} and LEP \cite{unknown:2004qh}.

An interesting conclusion of the full analysis is that the
present measurements of the anomalous magnetic moment define the
following preferred parameter region for the disformal model:
\begin{equation}
\text{6.0 GeV }\gtrsim \frac{f^{4}}{N^{1/2}\Lambda ^{3}}\gtrsim
\text{ 1.6 GeV (95\%\;c.l.)}\;,
\end{equation}
where we have updated the limits derived in Ref. \citen{BWRad} with the
current discrepancy between the measured and SM values
for the anomalous magnetic moment of the muon:
$\delta a_\mu=(26.1\pm 8.0)\times 10^{-10}$ \cite{muong2}.
As a consequence of this result and by taking into account the
prospects shown in Table \ref{radcoll}, the first signals of disformal
fields at colliders may be associated with radiative corrections \cite{BWRad}.

%%%%%%%%%%%%%%%%%%%%%%%%%%%%%%%%%%%%%%%%%%%
%% Table 2
%%%%%%%%%%%%%%%%%%%%%%%%%%%%%%%%%%%%%%%%%%%
\begin{table}
\tbl{
Summary of virtual disformal field searches at colliders (at $95\;\%$ c.l.).
$\gamma\gamma$, $e^+e^-$ and $e^+p$ ($e^-p$) channels are denoted by The upper indices ${a,b,c}$,
respectively. The prospects for the LHC are compared with current constraints from HERA, LEP and Tevatron  \cite{BWRad}.
The first two columns have the same information that in Table \ref{tabHad}. The third column shows the lower bound on
$f^2/(N^{1/4}\Lambda)$.}
{\begin{tabular}{||c|c c c||} \hline
%\hline
Experiment          & $\sqrt s$ (TeV) & ${\cal L}$ (pb$^{-1}$) &
$f^2/(N^{1/4}\Lambda)$ (GeV) \\ \hline
HERA$^{\,c}$        & 0.3             &  117                   & 52                           \\
Tevatron$^{\,a,\,b}$   & 1.8        &  127                   & 69                           \\
LEP$^{\,a}$      & 0.2             &  700                   & 59                           \\
LEP$^{\,b}$      & 0.2             &  700                   & 75
\\ \hline
LHC$^{\,b}$     & 14              & $10^5$                 & 383
\\ \hline
%\hline
\end{tabular}}\label{radcoll}
\end{table}
%%%%%%%%%%%%%%%%%%%%%%%%%%%%%%%%%%%%%%%%%%%
%% Table
%%%%%%%%%%%%%%%%%%%%%%%%%%%%%%%%%%%%%%%%%%%

\begin{figure}%[bt]
\begin{center}
%\resizebox{8.8cm}{6.4cm}
\resizebox{10cm}{!}
{\includegraphics{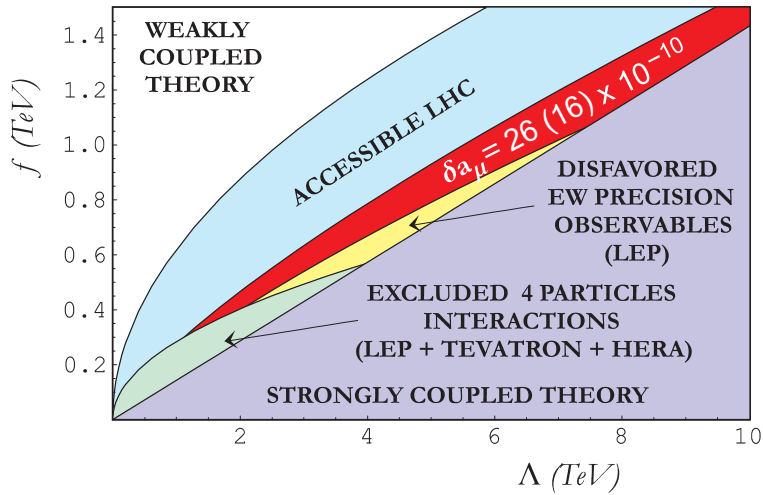}} \caption {Parameter space limits for disformal radiative corrections for a model
with N = 1. In the (red) central area, the disformal model can account for the muon magnetic moment deficit,
 being consistent with currente collider constraints (the most important one is associated with the the Bhabha scattering at LEP)
and electroweak precision observables. Prospects for the sensitivity of future colliders are also estimated.\cite{BWRad}
\label{FigRad}}
\end{center}
\end{figure}

\section{Relic abundances for disformal scalars}
\label{Rel}

We can calculate the thermal relic abundance associated to disformal scalars by
taking into account the disformal coupling given by Eq. (\ref{lag}) as it was
done in Refs. \citen{CDM,CDM2}. In this chapter, we will review the basic steps
of this computation by distinguishing
if the disformal field is relativistic (hot) or non-relativistic (cold) at decoupling.
The evolution of the number density $n_\alpha$ of the scalar $\pi^\alpha$,
 $\alpha=1,\dots , N$ with $N$ the number of different types of disformal fields,
is given by the Boltzmann equation:
\begin{eqnarray}
\frac{dn_\alpha}{dt}=-3Hn_\alpha-\langle \sigma_A v\rangle
(n_\alpha^2 -(n_\alpha^{eq})^2)\,.
\label{Boltzmann}
\end{eqnarray}
Here,
\begin{eqnarray}
\sigma_A=\sum_X \sigma(\pi^\alpha\pi^\alpha\rightarrow X)
\end{eqnarray}
is the total annihilation cross-section of $\pi^\alpha$ into SM particles $X$.
$H$ is the Hubble parameter, and the $-3Hn_\alpha$ term takes into account the
dilution of the number density due to the expansion of the universe.
Under the disformal parity symmetry discussed in the introduction, they do not decay into other
particles. Therefore, these are the only terms which could change the number density
of disformal scalars. For simplicity, we will assume that all the disformal scalars
have the same mass. It implies that each disformal species evolves in an independent way,
and we will remove the $\alpha$ super-index.

The thermal average $\langle \sigma_A v\rangle$ of the total annihilation cross-section times the relative velocity can be written as
\begin{eqnarray}
\langle \sigma_A v\rangle=\frac{1}{n_{eq}^2}\int
\frac{d^3p_1}{(2\pi)^3}
\frac{d^3p_2}{(2\pi)^3} f(E_1) f(E_2)\frac{w(s)}{E_1 E_2}\,.
\end{eqnarray}
Here,
\begin{eqnarray}
w(s)=E_1 E_2 \sigma_A v_{rel}=\frac{s\sigma_A}{2}
\sqrt{1-\frac{4 M^2}{s}}\,.
\end{eqnarray}
In this case, the Mandelstam variable $s$ has the standard definition in terms of the four-momenta of the two branons $p_1$ and
$p_2$ as $s=(p_1+p_2)^2=2(M^2+E_1E_2-\vert \vec p_1\vert \vert \vec p_2\vert \cos\theta)$. We can assume a vanishing chemical potential
for the disformal scalars, whose distribution functions are given by the Bose-Einstein one:
\begin{eqnarray}
f(E)=\frac{1}{e^{E/T} + a}\,,
\end{eqnarray}
with $a=-1$. In the non-relativistic case $T\ll 3M$, the Maxwell-Boltzmann distribution ($a=0$) is a good approximation, which we will use.
The equilibrium abundance is just:
\begin{eqnarray}
n_{eq}=\int \frac{d^3p}{(2\pi)^3} f(E)\,.
\end{eqnarray}
The previous thermal average includes annihilations into all the SM particle-antiparticle pairs. If the temperature of the thermal bath is above the
QCD phase transition ($T>T_c$), we consider  annihilations into quark-antiquark and gluons pairs. In the opposite case ($T<T_c$), we include annihilations into light hadrons. For numerical computations, we assume a critical temperature $T_c\simeq 170$ MeV although the final results are not very sensitive to the concrete value of this parameter.

To solve the Boltzmann equation is common to define $x=M/T$ and $Y=n/s_e$ with $s_e$ the entropy density of the entire thermal bath. We will assume that the total entropy is conserved, i.e. $S_e=a^3 s_e=\mbox{const}$, where $a$ is the scale factor.  On the other hand, the Friedmann equation reads
\begin{eqnarray}
H^2=\frac{8\pi}{3M_P^2}\rho\,.
\end{eqnarray}
%%%%%%%%%%%%
The energy density in a radiation dominated universe can be written as
\begin{eqnarray}
\rho=g_{eff}(T)\frac{\pi^2}{30}T^4\,,
\end{eqnarray}
whereas, the entropy density is given by
\begin{eqnarray}
s_e=h_{eff}(T)\frac{2\pi^2}{45}T^3\,.
\end{eqnarray}
Here, $g_{eff}(T)$ and $h_{eff}(T)$ mean the effective number of relativistic degrees of freedom contributing
to the energy density and to the entropy density respectively at a given temperature of the photon thermal bath $T$.
For $T\gtrsim 1$ MeV, $h_{eff}\simeq g_{eff}$. Using all these definitions:
\begin{eqnarray}
\frac{dY}{dx}=-\left(\frac{\pi M_P^2}{45}\right)^{1/2}
\frac{h_{eff}M}{g_{eff}^{1/2}x^2}\langle\sigma_A v
\rangle(Y^2-Y_{eq}^2)\,,
\label{YBoltz}
\end{eqnarray}
where we have neglected the contribution from derivative terms of the form $dh_{eff}/dT$. We can introduce the annihiliation rate as
$\Gamma_A=n_{eq}\langle\sigma_A v\rangle$. If $\Gamma_A$ is larger than the expansion rate of the universe $H$ at a given $x$, then
$Y(x)\simeq Y_{eq}(x)$, i.e., the abundance of disformal scalars is determined by the equilibrium one. On the contrary, as
$\Gamma_A$ decreases with the temperature, it eventually becomes similar to $H$  at some point $x=x_f$. From that
time on, disformal particles will be decoupled from the SM thermal bath and its abundance will remain frozen, i.e.
$Y(x)\simeq Y_{eq}(x_f)$ for $x\geq x_f$. For relativistic (hot) relics, the equilibrium abundance can be written as
\begin{eqnarray}
Y_{eq}(x)=\frac{45\zeta(3)}{2\pi^4}\frac{1}{h_{eff}(x)},
\;\;\;\;\; (x \ll 3)\,.\label{hot}
\end{eqnarray}
For non-relativistic (cold) particles:
\begin{eqnarray}
Y_{eq}(x)=\frac{45}{2\pi^4}\left(\frac{\pi}{8}
\right)^{1/2}x^{3/2}\frac{1}{h_{eff}(x)}\,e^{-x},
\;\;\;\;\; (x \gg 3)\,.\label{cold}
\end{eqnarray}
In the case of hot disformal fields, the equilibrium abundance is not very sensitive to the value of $x$. The situation is different for cold relics since $Y_{eq}$ decreases exponentially with the temperature. The sooner the decoupling takes place, the larger the relic abundance. First, we will discuss the relativistic decoupling. In such a case, the equilibrium abundance depends on $x_f$ only through $h_{eff}(x_f)$ and it is a good approximation to impose the condition $\Gamma_A=H$:
\begin{eqnarray}
H(T_f)=1.67\, g_{eff}^{1/2}(T_f) \frac{T_f^2}{M_P}=\Gamma_A(T_f)\,.
\label{Hubble}
\end{eqnarray}
It can be solved for $T_f$ by expanding $\Gamma_A(T_f)$ in the relativistic limit $T_f\gg M/3$.

Once $x_f$ is known, we can compute the current number density of disformal particles and its
corresponding energy density by taking into account Eq.(\ref{hot}) ($Y_{\infty}\simeq Y(x_f)$):
\begin{eqnarray}
\Omega_{D} h^2=7.83 \cdot 10^{-2} \frac{1}{h_{eff}(x_f)}
\frac{M}{\mbox{eV}}\,.
\label{eV}
\end{eqnarray}
For cold disformal fields, the computation of the decoupling temperature is a little more involved. We can use the well-known
semi-numerical result\cite{Kolb}:
\begin{eqnarray}
x_f=\ln\left(\frac{0.038\, c\,(c+2) M_P M \langle\sigma_A v
\rangle}{g_{eff}^{1/2}\,x_f^{1/2}}\right)\,,
\label{xf}
\end{eqnarray}
with $c\simeq 0.5$. The above equation can be solved iteratively with the result:
\begin{eqnarray}
\Omega_{D} h^2=8.77 \cdot 10^{-11} \mbox{GeV}^{-2}
\frac{x_f}{g_{eff}^{1/2}}\left(\sum_{n=0}^\infty
\frac{c_n}{n+1}x_f^{-n}\right)^{-1}\,.
\label{coldomega}
\end{eqnarray}
Here, we have explicitly used the standard expansion of $\langle \sigma_A v\rangle $ in powers of $x^{-1}$:
\begin{eqnarray}
\langle \sigma_A v\rangle =\sum_{n=0}^\infty c_n x^{-n}\,.
\end{eqnarray}
It means that the weaker the cross-section, the larger the relic abundance. This conclusion is expected since the sooner the decoupling takes place, the larger the relic abundance, and the decoupling occurs earlier for a weaker interaction. Therefore the cosmological bounds associated with the relic abundance are complementary to those coming from particle colliders. It implies that a constraint such as $\Omega_{D} < \Od(0.1)$ means
a lower limit for the value of the cross-sections in contrast with the upper limit we discussed in the previous section from non observation effects in particle collisions.

The above formalism can be applied to obtain the relic abundance of disformal fields $\Omega_{D} h^2$, both for the relativistic and non-relativistic decoupling. For that purpose, we can use the thermal averages $\langle \sigma_A v\rangle $ summarized in \ref{Cross} and \ref{Thermal} and computed in Refs. \citen{CDM,CDM2}. The disformal coupling depends only on the nature of the concrete SM particle. We will take into account the annihilation of disformal fields into photons, massive $W^{\pm}$ and $Z$ gauge bosons, three massless Majorana neutrinos, charged leptons, quarks and gluons (or light hadrons depending on the temperature of the thermal bath) and a real scalar Higgs field, in terms of the brane tension $f$ and the generic disformal mass $M$.

\section{Constraints on disformal dark matter abundance}
\label{RelCon}

\begin{figure}%[h]
\centerline{\psfig{file=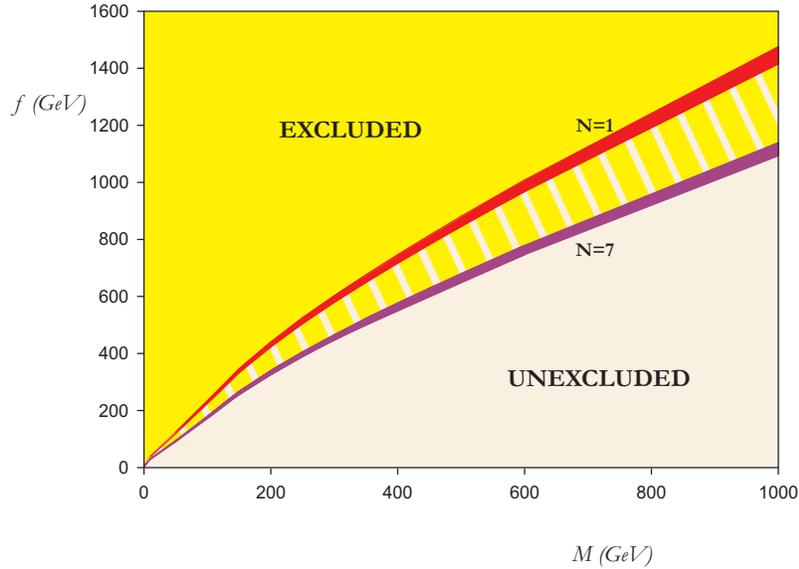,width=10.5cm}}
%\vspace*{8pt}
%{\epsfxsize=12.0 cm \epsfbox{coldlin2.eps}}
\caption{Exclusion limits for the abundance of disformal fields as cold DM.
The thick lines have associated an abundance of  $\Omega_{D}h^2=0.126 - 0.114$ \cite{Ade:2015xua}
for $N=1$ and $N=7$. In this sense, the region above the curves ($\Omega_{D}h^2>0.126$) is excluded.
The area in between both lines contains the models with $1<N<7$.\cite{CDM,CDM2}\label{coldplot}}
\end{figure}

We shall discuss first the case of cold disformal fields. By taking into account the $c_n$ coefficients for the total cross-section summarized in \ref{Thermal}, the computation of the freeze-out value $x_f$ from Equation (\ref{xf}) is straightforward \cite{CDM,CDM2}. It allows to
evaluate the relic abundance $\Omega_{D}h^2$ from Eq. (\ref{coldomega}) in terms of $f$ and $M$. We can impose the observational limit on the total
cold DM (CDM) density from Planck: $\Omega_{D}h^2 < 0.126 - 0.114$ at the 95$\%$ C.L. \cite{Ade:2015xua},
and the results can be observed in Fig. \ref{coldplot}. \cite{CDM,CDM2}.
The validity of the cold (hot) relic approximation can be found also in Fig. \ref{Combined}, where we have plotted the $x_f=3$ line.
The excluded region by this argument is located between the two curves. On the other hand, the solid line shows explicitly
the values of the disformal model parameters where they can constitute the entire DM density of our universe.

\begin{figure}%[h]
%{\epsfxsize=12.0 cm \epsfbox{AB1237WM2.eps}}
\centerline{\psfig{file=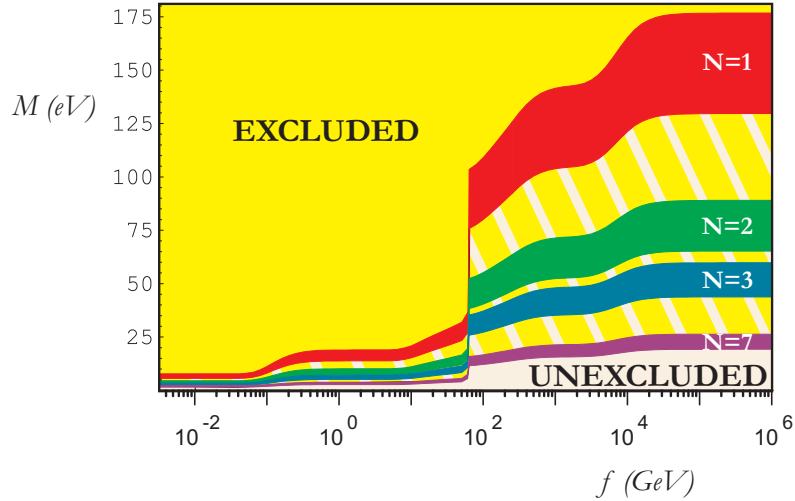,width=10.5cm}}
%\vspace*{8pt}
\caption{Exclusion limits for $N=1,2,3,7$ disformal scalars from the total DM abundance corresponding to hot relics.
The shaded regions is associated with the total DM limit $\Omega_{D}h^2=0.126 - 0.114$ \cite{Ade:2015xua}
for a given $N$. The area above these region is excluded by overproduction of disformal DM.\cite{CDM,CDM2}\label{Totalhot}}
\end{figure}

For hot disformal particles, we can find the freeze-out temperature $T_f$ in terms of the coupling scale $f$
by using Equation (\ref{Hubble}). Indeed, there is an approximated power law relation between both quantities:
$\log_{10} (f/1 \mbox{GeV}) \simeq(7/8)\log_{10}(T_f/1 \mbox{ GeV}) + 2.8$ \cite{CDM,CDM2}. This relation is not very
sensitive to the number of disformal scalars and allows to obtain $\Omega_{D}h^2$ from (\ref{eV}).
We can consider two types of bounds. On the one hand, the constraints associated with the total DM of the universe
$\Omega_{D}h^2=0.126 - 0.114$ \cite{Ade:2015xua} (Fig. \ref{Totalhot}). On the other hand, there are more
constraining bounds on the amount of hot DM since the free-streaming effect associated to this type of DM reduces
the power of structures on small scales, modifying the shape of the matter power spectrum.
The present limit reads $\Omega_{D}h^2<0.0071$ at the 95$\%$C.L. and it is obtained
from a combined analysis of data coming from BAO, JLA and Planck \cite{Ade:2015xua}.
These limits are plotted in Fig.\ref{hot}. The important increase for $f\simeq 60$ GeV in Figs. \ref{Totalhot}
and \ref{hot} (and also in Fig. \ref{Combined})
is related to a decoupling temperature of $T\simeq 170$ MeV, that is the one we have assumed for the QCD phase transition.
Thus, the jump is associated with a sudden growth in the number of  effective degrees of freedom
when entering in the quark-gluon plasma from the strong hadronic phase. The exact exclusion areas depend on the
number of disformal scalar fields. In Figs. \ref{Totalhot} and \ref{hot} is possible to see these limits for $N=1,2,3,7$.

The previous constraints assume that the disformal particles are relativistic
at  freeze-out. If we require $x_f \ll 3$, the limits are not valid for $f<10^{-4}$ GeV.
The line $x_f=3$ is explicitly plotted in Fig. \ref{Combined} in the hot relic approximation.

In addition to $f\ll M_D$, the above discussion assume a standard cosmological evolution
up to a temperature around $f$. Indeed, the effective action (\ref{lag}) is only valid at low energies
in relation to $f$. This scale fixes the range of validity of the results. We have checked that the
previous computations are consistent with these assumptions. In particular, the decoupling temperature
is always smaller than $f$ in the allowed regions of the different figures.

\section{Disformal dark matter direct detection}\label{direct}
\label{Dir}

If we assume that the DM halo of the Milky Way is composed of disformal fields,
its flux on the Earth is of order $10^{5}(100$ GeV$/M )$ cm$^{-2}$s$^{-1}$. This
flux could be sufficiently large to be measured in direct detection
experiments such as DAMA, EDELWEISS-II, CoGeNT, XENON100 or CDMS II.
These experiments measure the rate $R$, and energy
$E_{R}$ of nuclear recoils \cite{LHCDirect}.

The differential counting rate for a nucleus with mass $m_{N}$ is given by
\begin{equation}
\frac{dR}{dE_{d}}=\frac{\rho _{0}}{m_{N} M}\int_{v_{\min
}}^{\infty }vf\left( v\right) \frac{d\sigma _{BrN}}{dE_{R}}\left(
v,E_{R}\right) dv\,.
\end{equation}
Here, $\rho_{0}$ is the local disformal density, $(d\sigma
_{DN}/dE_{R})(v,E_{R})$ is the differential cross-section for the
disformal-nucleus elastic scattering and $f(v)$ is the speed
distribution of the disformal particles in the detector frame normalized to unity.

\begin{figure}%[h]
%{\epsfxsize=12.0 cm \epsfbox{AN1237WM2.eps}}
\centerline{\psfig{file=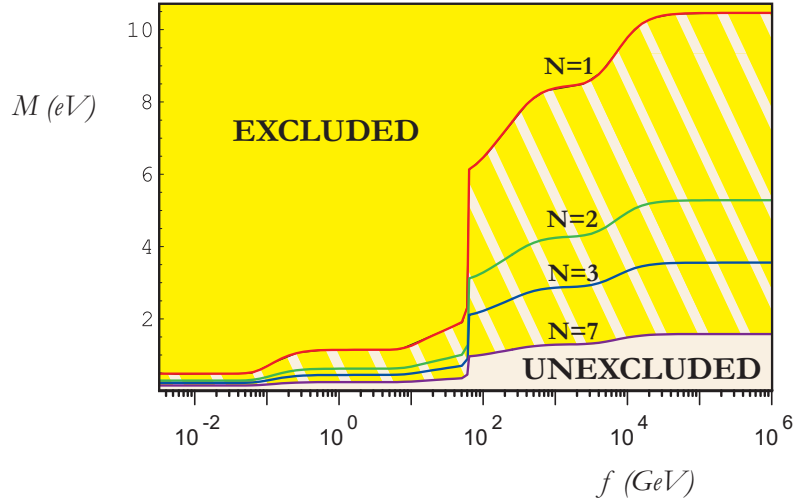,width=10.5cm}}
%\vspace*{8pt}
\caption{Exclusion limits for hot relics in models with $N=1,2,3,7$ disformal scalar particles.
For different $N$, the curves shows the $\Omega_{D}h^2=0.0071$ DM abundance. The area above
the curves is excluded depending on $N$.\cite{CDM,CDM2}\label{hot}}
\end{figure}

The relative speed of DM particles is of order 100 km s$^{-1}$, so the elastic scattering is non-relativistic. Therefore,
the recoil energy of the nucleon in terms of the scattering angle in
the center of mass frame $\theta^{*}$, and the disformal-nucleus
reduced mass $\mu_{N}=Mm_{N}/(M+m_{N})$, can be written as \cite{LHCDirect}
\begin{equation}
E_{R}=\frac{\mu _{N}^{2}v^{2}\left( 1-\cos \theta ^{\ast }\right)
}{m_{N}}\,.
\end{equation}
On the one hand, the lower limit of the integral is given in terms of the minimum
disformal particle speed, which is able to produce a recoil of energy
$E_{R}$: $v_{min}=(m_{N}E_{R}/2\mu_{N}^{2})^{1/2}$. On the other hand,
the upper limit is not bounded, but it is interesting to note that a
value of $v_{esc}=$ 650 km s$^{-1}$ is standard for the local escape speed
$v_{esc}$ of WIMPs gravitationally captured by the Milky Way \cite{cerdeno}.

The total event rate of collisions of disformal particles with matter
per kilogram per day $R$, can be computed by integrating the differential event rate
over all the possible recoil energies. The threshold energy $E_{T}$ is
the smallest recoil energy that the detector is able to measure. In terms of $E_{T}$,
this total rate can be written as \cite{LHCDirect}:
\begin{equation}
R=\int_{E_{T}}^{\infty }dE_{R}\frac{\rho _{0}}{m_{N}M}
\int_{v_{\min }}^{\infty }vf\left( v\right) \frac{d\sigma
_{Br N}}{dE_{R}}\left( v,E_{R}\right) dv\,.
\end{equation}
The disformal-nucleus differential cross-section is determined by
the interaction described by Eq. (\ref{lag}). For a general DM candidate, its interaction with nucleus
is separated into a spin-independent (SI) and a spin-dependent (SD) contribution \cite{LHCDirect}:
\begin{equation}
\frac{d\sigma _{N}}{dE_{R}}=\frac{m_{N}}{2\mu _{N}^{2}v^{2}}\left(
\sigma _{0}^{SI}F_{SI}^{2}\left( E_{R}\right) +\sigma
_{0}^{SD}F_{SD}^{2}\left( E_{R}\right) \right)\,,
\end{equation}
where the form factors $F_{SI}(E_{R})$ and $F_{SD}(E_{R})$ account
for coherence effects. It includes a suppression in the event
rate for heavy WIMPs on nucleons and gives the dependence on the
momentum transfer $q=\sqrt{2m_{N}E_{R}}$. The spin-independent and spin-dependent cross-sections
at zero momentum transfer are $\sigma_{0}^{SI}$ and $\sigma_{0}^{SD}$ respectively.
These quantities depend on the nuclear structure and the isospin content, i.e.
the number of protons and neutrons \cite{kopp}. In the case of disformal particles,
the entire interaction is SI and is given by \cite{kopp}
\begin{equation}
\sigma^{\rm SI}=\frac{[Zf_p+(A-Z)f_n]^2}{f_p^2}\frac{\mu_{D
n}^2}{\mu_{D p}^2}\sigma_p^{\rm SI}\,,
\end{equation}
where $Z$ is the charge of the nucleus, $A$ is the atomic mass number,
$f_{p,n}$ is the SI DM coupling to proton and neutron respectively,
$\mu_{D p}$ is the reduced mass of the disformal particle-proton system,
and $\sigma_p^{SI}$ is the SI cross-section for scattering of the disformal
particle on a proton. In particular, disformal fields do not violate the isospin
symmetry if we neglect the proton-neutron mass difference. Within this
approximation: $f_{p}=f_{n}$, and $\mu\equiv\mu_{D n}=\mu_{D p}$.
Indeed, the disformal-nucleon cross-section $\sigma_{n}$ was computed in
Refs. \citen{CDM} and \citen{CDM2}:
\begin{equation}
\sigma_p^{\rm SI}=\sigma _{n}=\frac{9M^{2}{m_n}^{2}\mu ^{2}}{64\pi f^{8}}\,.
\end{equation}
Here, $m_n$ is the nucleon mass. Different direct search experiments
have reported possible candidate events associated with DM detection.
The DAMA/NaI and DAMA/LIBRA detectors at the Gran Sasso National Laboratory
found an annual modulation signature consistent with light WIMPs \cite{dama}.
Similar conclusions can be derived from data from the ultralow-noise germanium detector operated deep underground in
Soudan Underground Laboratory (CoGeNT) \cite{CoGeNT}. However, these measurements are in clear tension with constraints
of other experiments. For example, from detectors located in the same laboratories such as XENON100
(a liquid xenon detector at the Gran Sasso National Laboratory) \cite{xenon}, or CDMS II (a germanium and silicon detector at the
the Soudan Underground Laboratory) \cite{CDMSevents}.

\begin{figure}
\begin{center}\includegraphics[width=9cm]{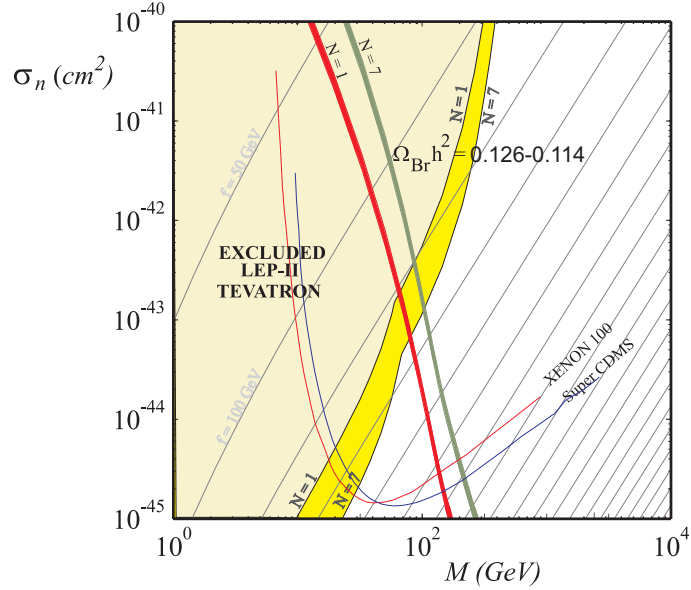} \end{center}
\caption{Elastic nuclear cross-section for disformal scalars $\sigma_{n}$
in terms of its mass $M$. The two thick lines have associated the
$\Omega_{D}h^{2}=0.126 - 0.114$ abundances for cold disformal relics with $N=1$
(left) and $N=7$ (right). The shaded regions on the left are
excluded by colliders \cite{BW2,Cembranos:2004jp} for $N=1$ and $N=7$ respectively.
Current limits from direct searches are shown with solid lines.\cite{CDM,LHCDirect}\label{figJose}}
\end{figure}

The results for direct searches of disformal DM in terms of its mass
and the restrictions from present experiments are shown in Figure \ref{figJose}.
Curves of constant $f$ with 50 GeV separation are shown for reference. The regions on
the left of the $\Omega_{D}h^{2}=0.126 - 0.114$ lines are excluded
by overproduction of disformal particles. On the contrary, the right areas are compatible with
current observations. Such regions correspond to $f\gtrsim 120$ GeV and
$M\gtrsim 40$ GeV.

\section{Indirect searches of disformal dark matter}
\label{Ind}

If DM is formed by disformal particles, two of them can annihilate into ordinary particles such as leptons, quarks and gauge bosons. Their annihilations from different astrophysical origins (galactic halo, Sun, Earth, etc.) produce fluxes of cosmic rays. Depending on the features of
these fluxes, they may be discriminated from the background through distinctive signatures. After the annihilation, the particle species that can be potentially detected by different experimental devices would be gamma rays, neutrinos and antimatter (fundamentally, positrons and antiprotons). In particular, gamma rays and neutrinos have the advantage of maintaining their original trajectory. On the contrary, analyses of charged cosmic rays are more involved due to galactic diffusion.

For example, the differential gamma-ray flux from annihilating DM particles is given, in general, by \cite{Vivi1}:
\begin{eqnarray}
\frac{\text{d}\,\Phi_{\gamma}^{\text{DM}}}{\text{d}\,E_{\gamma}} =
\frac{1}{4\pi M^2}\sum_i\langle\sigma_i v\rangle
\frac{\text{d}\,N_\gamma^i}{\text{d}\,E_{\gamma}}\, \times\, \frac{1}{\Delta\Omega}\int_{\Delta\Omega}\text{d}\Omega\int_{l.o.s.} \rho^2[(s)] \text{d}s\,.
\label{flux}
\end{eqnarray}
Here, the second term on the r.h.s. of this equation is commonly known as the astrophysical factor, which is defined by the integral of the DM mass density profile $\rho(r)$, along the line of sight ($l.o.s.$) between the source and the detector (divided by the detector solid angle). On the contrary, the first term depends on the nature of the DM particle, since $\langle\sigma_i v\rangle$ is the averaged annihilation cross-section times velocity of two DM particles into two SM particles (of a given type, labeled by the subindex $i$). In order to compute
the number of photons produced in each annihilating channel per energy $\text{d}\,N_\gamma^i/\text{d}\,E_{\gamma}$, it is necessary to take into account different SM decays and/or hadronization of unstable produced particles such as gauge bosons and quarks. Due to the non-perturbative QCD effects, the study of these particle chains requires Monte Carlo events generators such as PYTHIA \cite{PYTHIA}.  In Ref. \citen{Ce10}, it was shown that the photon spectra for the SM particle-antiparticle channels can be written in terms of three different parametrizations. For leptons and light quarks, excluding the top, one can write:
\begin{eqnarray}
x^{1.5}\frac{\text{d}N_{\gamma}}{\text{d}x}\,&=&\, a_{1}\text{exp}\left(-b_{1} x^{n_1}-b_2 x^{n_2} -\frac{c_{1}}{x^{d_1}}+\frac{c_2}{x^{d_2}}\right) \nonumber\\
&+& q\,x^{1.5}\,\text{ln}\left[p(1-x)\right]\frac{x^2-2x+2}{x}\,;
\label{general_formula}
\end{eqnarray}
whereas for the top quark, the expression reads:
\begin{eqnarray}
x^{1.5}\frac{\text{d}N_{\gamma}}{\text{d}x}\,=\, a_{1}\,\text{exp}\left(-b_{1}\, x^{n_1}-\frac{c_{1}}{x^{d_1}}-\frac{c_{2}}{x^{d_2}}\right)\left\{\frac{\text{ln}[p(1-x^{l})]}{\text{ln}\,p}\right\}^{q}\,.
\label{general_formula_t}
\end{eqnarray}
On the other hand, for the electroweak gauge bosons ($W$ and $Z$):
\begin{eqnarray}
x^{1.5}\frac{\text{d}N_{\gamma}}{\text{d}x}\,=\, a_{1}\,\text{exp}\left(-b_{1}\, x^{n_1}-\frac{c_{1}}{x^{d_1}}\right)\left\{\frac{\text{ln}[p(j-x)]}{\text{ln}\,p}\right\}^{q}\,.
\label{general_formula_W_Z}
\end{eqnarray}
In these expressions, $x\equiv E_{\gamma}/M$ where $E_{\gamma}$ holds for the photon energy and $M$ is the mass of the disformal particle.
The particular values of the constants in the above equations can be found online \cite{Online} and in the original manuscript \cite{Ce10}.
For each channel, some of these parameters depend on $M$, whereas others are constant.

\begin{figure}[bt]
\begin{center}
%\epsfxsize=9cm
%\resizebox{11cm}{8cm}
\resizebox{10cm}{!}
{\includegraphics{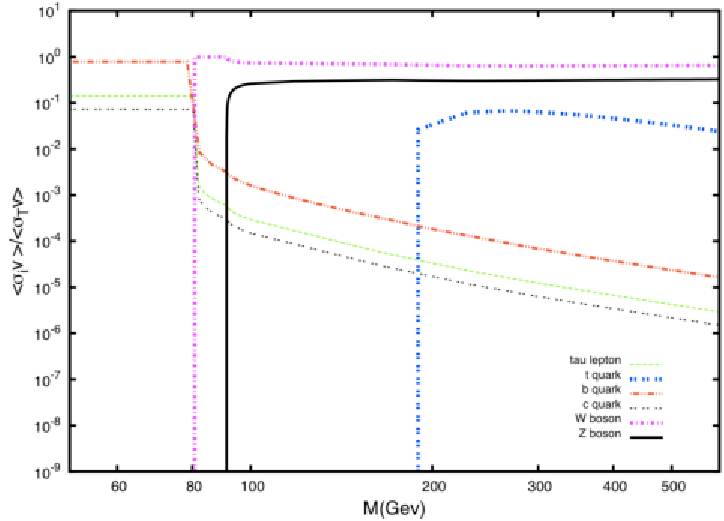}}
%{\includegraphics{Fig11.pdf}}
\caption {Branching ratios for disformal particles annihilatining into SM particle-antipaticle pairs. For heavy disformal scalars, the main contribution to the photon flux comes from the gauge bosons $W^+W^-$ and $ZZ$ annihilating channels. On the contrary, if
these channels are kinematically forbidden, the rest of contributions have to be computed.\cite{Vivi1}\label{BR}}
\end{center}
\end{figure}

Finally, in order to calculate the gamma-ray spectra, we need to compute the total annihilation cross-section and its annihilation branching ratios into SM particle-antiparticle pairs. The annihilation cross-sections for disformal particles only depend on the spin and mass of the produced SM pairs. They can be found in Appendix C \cite{CDM,CDM2}. For instance, the leading term for non-relativistic disformal fields annihilating into a Dirac fermion $\psi$ with mass $m_\psi$ can be written as:
\begin{eqnarray}
\langle \sigma_{\psi} v\rangle\,=\,\frac{1}{16\pi^2f^8}M^2 m_\psi^2\left(M^2-m_\psi^2\right)\,\sqrt{1-\frac{m_\psi^2}{M^2}}\,;
\end{eqnarray}
for a massive gauge boson $Z$ (with mass $m_Z$):
\begin{eqnarray}
\langle \sigma_{Z} v\rangle\,=\,
\frac{1}{64\pi^2f^8}
M^2\,\left( 4\,M^4 - 4\,M^2\,{m_Z}^2 + 3\,
{m_Z}^4 \right)\,{\sqrt{1 - \frac{{m_Z}^2}{M^2}}}
\,;
\end{eqnarray}
whereas for a massless gauge field $\gamma$, the leading order is zero:
\begin{eqnarray}
\langle \sigma_{\gamma} v\rangle&=&0;
\end{eqnarray}
and, finally, for a (complex) scalar field $\Phi$ of mass $m_\Phi$:
\begin{eqnarray}
\langle \sigma_{\Phi} v\rangle\,=\,\frac{1}{32\pi^2f^8}
M^2\,{\left( 2\,M^2 + {m_\Phi}^2 \right) }^2\,
{\sqrt{1 - \frac{{m_\Phi}^2}{M^2}}}
\,.
\end{eqnarray}
It is interesting to note that disformal particles produce gamma-ray lines from direct annihilation into photons.
However, this annihilation is highly suppressed since it takes place in $d$-wave channel. Therefore, we will not search for the
monochromatic signal at an energy equal to the disformal mass. In the case of heavy disformal particles, the
leading annihilation channels are into $W^+ W^-$  and $ZZ$ (Fig. \ref{BR}) \cite{CDM,Vivi1}. In this case, the energy of the produced gamma rays could be in the range 30 GeV-10 TeV. These high-energy photons may be detectable by Atmospheric Cherenkov Telescopes (ACTs) such as MAGIC \cite{Mag, Mag11}. On the other hand, if the annihilation into electroweak gauge bosons is kinematically forbidden ($M<m_{W,Z}$), the leading annihilation channel is into the heaviest kinematically allowed quark-antiquark pair (Fig. \ref{BR}). For this case of light disformal DM, photon fluxes are more suitable to be detected by space-based gamma-ray observatories \cite{CDM,Vivi1,Ce10,indirect} such as EGRET \cite{EGRET} and FERMI \cite{Fer, FER}, with better sensitivities for an energy range between 30 MeV and 300 GeV.
%
%%%%%%%%%%%%%%%%%%%%

\begin{figure}[t]
\begin{center}
%\epsfxsize=10cm   %width of figure - will enlarge/reduce the figures
%\resizebox{10.8cm}{8.4cm}
\resizebox{10cm}{!}
{\includegraphics{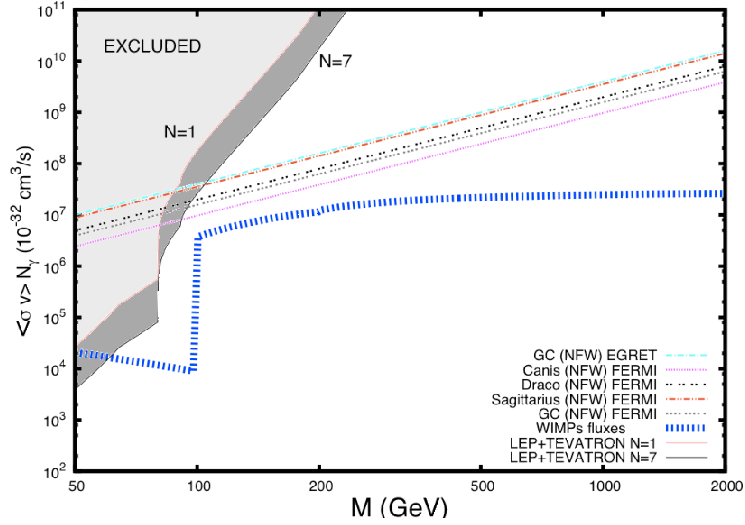}}
\caption{Sensitivity and constraints on disformal DM annihilation into gamma rays coming from different targets.
Exclusion limits (at $5\sigma$ ) for satellite experiments (FERMI and EGRET) are shown with
straight lines. The (blue) thick line is associated to the fluxes assuming a DM abundance
$\Omega_{\text{CDM}} h^2 = 0.126 - 0.114$ \cite{Ade:2015xua}. The upper left corner is
excluded by collider searches, and figure shows the limit from  $\text{LEP}$ and $\text{TEVATRON}$ experiments
for $N = 1$ and $N=7$ number of disformal particles.\cite{Vivi1}\label{FER}}
\end{center}
\end{figure}

\begin{figure}[t]
\begin{center}
%\epsfxsize=10cm   %width of figure - will enlarge/reduce the figures
%\resizebox{10.8cm}{8.4cm}
\resizebox{10cm}{!}
{\includegraphics{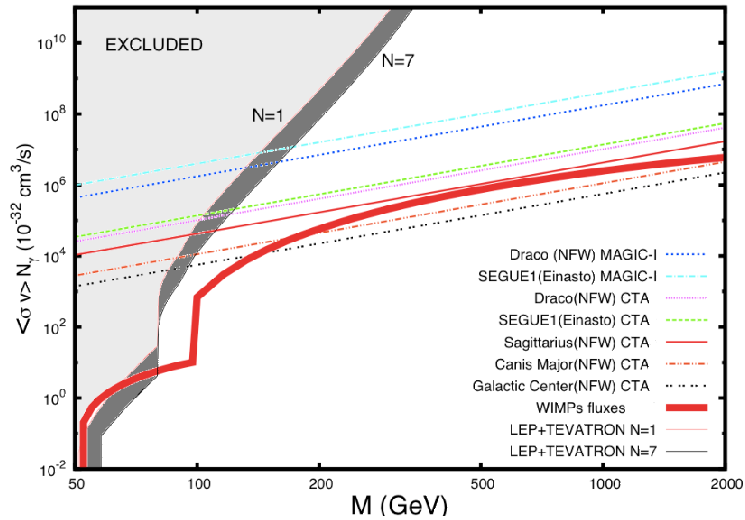}}
\caption {Similar to Fig. \ref{FER} but for current and future ground-based detectors.\cite{Vivi1}\label{ACT}}
\end{center}
\end{figure}

With the above information, it is possible to estimate the current constraints and sensitivity for different targets and
detectors as shown in Figs. \ref{FER} and \ref{ACT} \cite{Vivi1}.
The resulting constraints on the total number of gamma-rays  $N_\gamma \langle\sigma v\rangle$
does not depend on the number of disformal species since the proportional lower flux coming from the annihilation of a larger number of
disformal fields, is compensated by the associated higher abundance that a larger number of disformal species provides
(for a fixed coupling scale $f$).

By assuming the Planck relic density for disformal DM \cite{Ade:2015xua} and the standard astrophysical factors for different astrophysical sources such as dwarf galaxies or the galactic center, it is possible to obtain the straight lines in Figs. \ref{FER} and \ref{ACT},
which represent present constraints or expected sensitivity at $5\sigma$ for a particular target and detector. Fundamentally, present experiments (EGRET, FERMI or MAGIC) are unable to
detect signals from disformal DM annihilation. However, future experiments such as CTA may be able to detect gamma rays coming from such annihilations for disformal masses higher than 150 GeV for observations of the Galactic Center or Canis Major depending on the properties of the
DM density distribution.

The above analysis about indirect searches of disformal DM are based on particular assumptions about DM distributions \cite{Cembranos:2005us}.
In particular, they assume profiles that are in good agreement with non-baryonic cold DM simulations, such as the standard NFW profile \cite{Navarro:1996gj}. However, when the baryonic gas is taken into account, it can modify the gravitational potential increasing the DM density
in the center of the halo. This conclusion has been reached in different studies \cite{Blumenthal, Prada:2004pi}, although there are
discrepancies \cite{Romano}. If this fact is correct, there are two important consequences for gamma-ray detection. On the one hand, the central accessible region is reduced to few tenths of a degree; on the other hand, the fluxes are enhanced by several orders of magnitude \cite{Prada:2004pi}. Indeed, the interpretation of very high energy (VHE) gamma rays from the Galactic Center (GC) is in good agreement with
disformal DM annihilation from these types of compressed dark halos \cite{ViviGC}. These fluxes have been observed by several collaborations, such as CANGAROO \cite{CANG}, VERITAS \cite{VER}, MAGIC \cite{MAG} or Fermi-LAT  \cite{Vitale, ferm}, but we will discuss particularly on the data collected by the HESS collaboration from 2004 to 2006 relative to the HESS J1745-290 source \cite{Aha, HESS}.

\begin{table}
%\begin{center}
\tbl{Best fit parameters for the HESS J1745-290 data as gamma rays produced by disformal DM annihilation.
The dominant channels contributing are $W^+W^-$ and $ZZ$. The four fitted parameters can be found in this table.
The value of the disformal mass (TeV), the normalization factor of the signal
$C^2\equiv {\Delta\Omega\, \langle J_{(a)} \rangle_{\Delta\Omega}}/{(8\pi M^2)}$,
the normalitazion factor of the gamma-ray background $B\,(10^{-4}\, \text{GeV}^{-1/2} \text{cm}^{-1} \text{s}^{-1/2})$,
and its associated spectral index $\Gamma$. We also show the $\chi^2$ per degree of freedom (dof), and the astrophysical
factor in units of the one associated with a NFW profile: $b\equiv \langle J_{(2)} \rangle/\langle J^{\text{NFW}}_{(2)} \rangle$,
where $\langle J^{\text{NFW}}_{(2)} \rangle\simeq 280 \cdot 10^{23}\; \text{GeV}^2 \text{cm}^{-5}$.
$b$ is computed with the model fitted cross-section to Planck data
$\Omega_{\text{D}} h^2 = 0.126 - 0.114$ \cite{Ade:2015xua}:
$\langle \sigma v \rangle = (1.14\pm0.19)\cdot10^{-26}\; \text{cm}^{3} \text{s}^{-1}$.\cite{ViviGC}}
{\begin{tabular}{|c|c|c|}
\hline
\hline
 $M$ (TeV) & $C\; (10^{-2}\, \text{GeV}\, \text{cm}^{-1} \text{s}^{-1/2})$ & $B\; (10^{-4}\, \text{GeV}^{-1/2} \text{cm}^{-1} \text{s}^{-1/2})$  \\
\hline
$50.6 \pm 4.5$  & $1.57 \pm 0.13$                                          &$5.27\pm2.32$     \\
\hline
\hline
 $\Gamma$           & $\chi^2/\,$dof   & $b$              \\
\hline
 $2.80 \pm 0.15$   &  0.84            & $4843\pm1134$    \\
\hline
\hline
\end{tabular}}\label{DMBra}
%\end{center}
\end{table}

The absence of temporal variability of these VHE events suggests a different emission mechanism than the IR and X-ray emission \cite{X}.
The source is very localized, in a region of few tenths of degree around the GC. On the other hand, the spectrum is characterized by
a cut-off at several tens of TeVs \cite{HESS}. The origin of these gamma rays is not clear. This flux may have been produced by particle
propagation close to the supermassive black hole Sgr A and the Sgr A East supernova remnant \cite{ferm,SgrA}, but the spectral features of
the data are perfectly consistent with the photons produced by the annihilation of disformal DM particles \cite{ViviGC} if it is complemented by
a background, which is well motivated by the radiative effects originated by  particle acceleration in the vicinity of Sgr A East supernova and the supermassive black hole, as we have commented. Therefore, we can write the total flux as\cite{ViviGC}
\begin{equation}
\frac{d\Phi_{\text{Tot}}}{dE}=\frac{d\Phi_{\text{Bg}}}{dE}+\frac{d\Phi_{\text{DM}}}{dE}\,,
\label{gen}
\end{equation}
where we will assume a simple power-law
\begin{eqnarray}
\label{powerlaw}
\frac{d\Phi_{\text{Bg}}}{dE}=B^2 \cdot \left(\frac{E}{\mbox{GeV}}\right)^{-\Gamma}\;,
\end{eqnarray}
for the discussed background. This shape is motivated by the observations of  the source
IFGL J1745.6-2900 by Fermi-LAT, that is spatially consistent with the HESS J1745-290 source \cite{Cohen,ferm}.
The background parameters, $B$ and $\Gamma$, can be also fitted from HESS data together with the disformal
DM mass $M$ and the astrophysical factor. We take into account a perfect efficiency and a experimental energy resolution
of $15\%$ ($\Delta E/E\simeq0.15$) \cite{ViviGC}. The disformal mass needed for fitting the data is around 50 TeV.
For this range of masses, as we have commented, the main contribution comes from the $ZZ$ and $W^+W^-$ annihilation channels,
producing a similar amount of $Z$, $W^+$ and $W^-$ bosons since
 $\langle\sigma_{W^+W^-} v\rangle \simeq 2\langle\sigma_{ZZ} v\rangle \simeq M^6/(8\pi^2\, f^8)$
(we are assuming only one disformal species).
As soon as we know $M$, we can compute the coupling corresponding to the DM abundance in agreement with Planck data
\cite{Ade:2015xua}: $f=27.5 \pm 2.4$ TeV. Then, we can calculate the thermal averaged cross-section: $\langle \sigma v \rangle = \sum_{i=W,Z} \langle \sigma_i v\rangle= (1.14\pm0.19)\cdot10^{-26}\; \text{cm}^{3} \text{s}^{-1}$, and finally the astrophysical factor. In Table \ref{DMBra},
$\langle J_{(2)} \rangle$ is presented in units of the astrophysical factor associated with a standard NFW profile:
$b\equiv \langle J_{(2)} \rangle/\langle J^{\text{NFW}}_{(2)} \rangle$. The large mass of the disformal field in order to explain these data,
makes difficult to check its DM origin with collider or direct DM experiments. However, the study of other cosmic rays could be able to prove
or disprove the model \cite{ViviGCother}.

\begin{figure}[]
\begin{center}
\resizebox%{11cm}{8cm}
{9cm}{!}
{\includegraphics{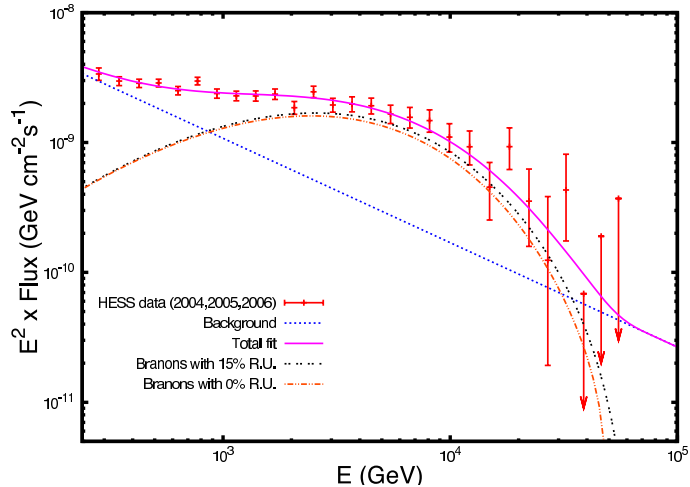}}
\caption {The total fit for the HESS J1745-290 data as gamma rays produced by disformal DM annihilation ($\chi^2/dof=0.84$).
The power law background with spectral index $\Gamma=2.80\pm0.15$ is shown with dotted line. The double-dotted line shows the
signal contribution with a $15\%$ of resolution uncertainity (R. U.) and normalization parameter $C=(1.57\pm0.24)\cdot10^{-7}\text{cm}^{-1}\text{s}^{-1/2}$.\cite{ViviGC}\label{BraGC}}
\end{center}
\end{figure}

\section{Nucleosynthesis constraints}
\label{BBNs}

On the other hand, there are astrophysical observations that can
restrict the parameter space of disformal scalars independently of
their abundance. For example,
one of the most successful predictions of the standard cosmological
model is the relative abundances of primordial elements. These abundances
are sensitive to several cosmological parameters and were used in Refs. \citen{CDM}
and \citen{CDM2}
in order to constrain the number of light branons. For instance, the production
of  $^4$He increases with an increasing rate of the expansion $H$. From (\ref{Hubble}),
we can deduce that the Hubble parameter depends on the effective number of relativistic degrees of
freedom $g_{eff}$. Traditionally, this number has been parametrized in terms
of the effective number of neutrino species $N_\nu=3+\Delta N_\nu$ in the following way:
\begin{eqnarray}
g_{eff}(T\sim \mbox{MeV})= g_{eff}^{SM}+g_{eff}^{new}\leq
10.75+\frac{7}{4}\Delta N_\nu\,.
\label{neutrino}
\end{eqnarray}
Here $T\sim \mbox{MeV}$ means the thermal bath temperature
at nucleosynthesis. In the SM, $g_{eff}^{SM}(T\sim \mbox{MeV})\simeq 10.75$
corresponds to the degrees of freedom associated with the photon, the electron,
and the three neutrinos.

Recent Planck data (combined with BAO) implies $\Delta N_\nu < 0.71$ at the 95$\%$C.L \cite{Ade:2015xua}. If we include the contribution
of disformal particles, the number of relativistic degrees of freedom at a given temperature $T$ can be written as:
\begin{eqnarray}
g_{eff}(T)=g_{eff}^{SM}(T)+N\left(\frac{T_D}{T}\right)^4\,.
\label{geff}
\end{eqnarray}
%%%%%%%%%%
Here $T_D$ is the  temperature of the cosmic disformal brackground,
whereas $g_{eff}^{SM}(T)$ denotes the contribution from the SM particles.
We are assuming that there are no additional new particles.
If the disformal scalars are not decoupled at a given temperature $T$,
they share the same temperature as the photons: $T_D=T$. On the contrary,
if they are decoupled, its temperature will be in general lower
than the one of the radiation. We can compute it by assuming
that the expansion is adiabatic:
\begin{eqnarray}
h_{eff}(T)=h_{eff}^{SM}(T)+N\left(\frac{T_D}{T}\right)^3\,,
\end{eqnarray}
where $h_{eff}^{SM}(T)$ takes into account the SM contribution.
If at some time between the freeze-out of disformal fiels and nucleosynthesis,
some other particle species become non-relativistic while still
in thermal equilibrium with the photon background,
its entropy is transferred to the photons, but not
to the disformal particles which are already decoupled. Therefore,
the entropy transfer increases the SM bath temperature relative
to the disformal temperature. If the total entropy of particles in equilibrium with
the photons remains constant:
\begin{eqnarray}
h_{eff}^{eq} a^3 T^3=\mbox{constant}\,.
\end{eqnarray}
On the other hand, since the number of relativistic
degrees of freedom $h_{eff}^{eq}$ has decreased, then
$T$ should increase with respect to $T_D$.
Thus, we find:
\begin{eqnarray}
\frac{g_{eff}^{eq}(T_{f,\,D})}{g_{eff}^{eq}(T)}=\frac{T^3}{T_D^3}\,,
\end{eqnarray}
where $T_{f,\,D}$ is the disformal freeze-out temperature, and
for particles in equilibrium with the photons
$g_{eff}^{eq}=h_{eff}^{eq}$. The final constraint on the number of
massless disformal species $N$ can be set by using (\ref{geff}):
\begin{eqnarray}
\frac{7}{4}\Delta N_\nu\geq N\left(\frac{T_D}{T_{nuc}}\right)^4=
N\left(\frac{g_{eff}^{eq}(T_{nuc})}
{g_{eff}^{eq}(T_{f,\,D})}\right)^{4/3}\,.
\end{eqnarray}
If the disformal particles decouple after nucleosynthesis,
the bound can be written as:
\begin{eqnarray}
N\leq \frac{7}{4}\Delta N_\nu \label{N1}\,.
\end{eqnarray}
On the contrary, if they decouple before, we have $g_{eff}^{eq}(T_{nuc})=10.75$,
and we can rewrite the limit as:
\begin{eqnarray}
N\leq \frac{7}{4}\Delta N_\nu\left(\frac{g_{eff}(T_{f,\,D})}{10.75}\right)^{4/3}\,.
\label{N2}
\end{eqnarray}
%%%%%%%%%%%%%%%%%%%

By taking $\Delta N_\nu=0.71$, we can find a relation between the disformal freeze-out temperature $T_{f,\, D}$ and
the coupling scale $f$ that constrains the number of disformal fields $N$ (see Fig. 4).
For $f< 10$ GeV, the restrictions are really important: $N\leq 1$. This conclusion is derived by using Eq.(\ref{N1})
for $f<3$ GeV (corresponding to $T_{f,D}\lesssim 1$ MeV) or Eq. (\ref{N2}) otherwise. However the bounds are less restrictive
in the range $f\simeq 10-60$ GeV: $N \leq 3$. On the other hand, above the QCD phase transition, $f\simeq 60$ GeV,
the limit increases so much that the restrictions become extremely weak. In this case, we are assuming exclusively the SM
content: $g_{eff}^{SM}(T\gtrsim 300 \;\mbox{GeV})=106.75$.
%%%%%%%%%%%

%
\begin{figure}%[h]
%{\epsfxsize=12.0 cm \epsfbox{Nuclecd1.eps}}
\centerline{\psfig{file=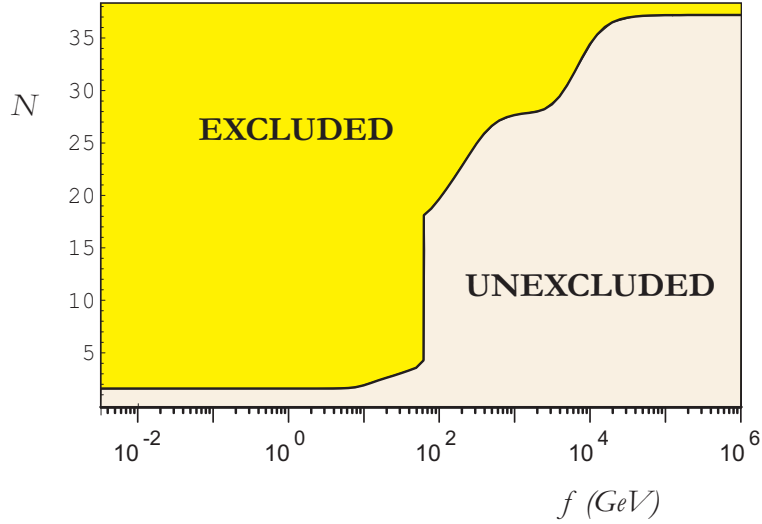,width=10.0cm}}
%\vspace*{8pt}
\caption{Constraints on the number of massless disformal scalar fields $N$
from nucleosynthesis as a function of the disformal coupling $f$ \cite{CDM,CDM2}.\label{BBN}
}
\end{figure}

\section{Constraints from supernova SN1987A}
\label{Ste}

Astrophysical bounds on the coupling scale can be obtained from modifications of cooling
processes in stellar objects like supernovae \cite{Kugo,CDM,Brax:2014vva}. These processes take place by
energy loosing through light particles such as photons and neutrinos. However, if the
disformal mass is low enough, disformal particles are expected to carry a fraction of this energy,
depending on their coupling to the SM fields. In order to analyze the constraints on $f$ and $M$
imposed by the cooling process of the neutron star in supernovae explosions, we will estimate
the energy emission rate from the supernova core by studying the electron-positron pair annihilation.
Disformal scalars produced within the core can be scattered or absorbed again depending on the strength
of their interaction with SM particles. The disformal mean free path $L$ inside the neutron star
needs to be larger than the star size ($R\sim \Od (10)$ Km) to be able to escape and contribute to
the cooling process. For a disformal scalar heavier than the supernova temperature $M\gg T_{SN}$,
we can estimate $L\sim (8\pi f^8)/(M^2 T_{SN}^4 n_e)$, where $n_e$ is the electron number density inside the star.
It means that the restrictions will apply only for $f\gtrsim 5$ GeV. The constraints arise because
the emitted energy in the form of disformal scalars could spoil the good agreement between the predictions for the
neutrino fluxes from supernova 1987A and the observations in Kamiokande II
\cite{kamiokandeII} and IMB \cite{IMB} detectors. The disformal emission could shorten
the duration of the neutrino signal with respect to the observed one, if the energy loss rate per
unit time and volume is $Q \gtrsim 5\times 10^{-30}$ GeV$^5$. For instance, the contribution of the mentioned electron-positron channel to the
volume emissivity can be written as \cite{CDM2}
\begin{eqnarray}
Q_{D}(f,M)&\equiv& \int\prod_{i=1}^2 \left\{\frac{d^3k_i}{(2\pi)^3
2E_i}2f_i \right\} (E_1+E_2)2s\,\,\sigma_{e^+e^-\rightarrow
\pi\pi}(s,f,M)\,.
\end{eqnarray}
Here, $i$ denotes the electron (1) and positron (2) particles (with negligible mass inside the supernova core).
The number density of electrons: $n_e\sim 1.4\times 10^{-3}\; \mbox{GeV}^3$, can be used to estimate the chemical
potential $\mu\sim (3 \pi^2 n_e)^{1/3}$, within the Fermi-Dirac distribution function $f_i=1/(e^{(E_i/T-\mu/T)}+1)$.
In such a case, $Q_{D}$ can be computed as \cite{CDM2}
\begin{eqnarray}
Q_{D}&=& \int_0^\infty dE_1\int_{M^2/E_1}^\infty dE_2\int_{-1}^{1
- 2 M^2/(E_1E_2)} d(cos)(E_1+E_2)
\nonumber\\
&&\frac{N[2E_1E_2\left(2E_1E_2(1-\cos)-4M^2\right)]^{5/2}\left(1-\cos\right)^{3/2}}
{{(2\pi)}^5\,7680\,f^8\,\left( 1 + e^{\frac{E_1-\mu}{T}} \right)
    \,\left( 1 + e^{\frac{E_2+\mu}{T}} \right)}\,.
\label{Q}
\end{eqnarray}

\begin{figure}%[h]
%{\epsfxsize=12.0 cm \epsfbox{Tres3.eps}}
\centerline{\psfig{file=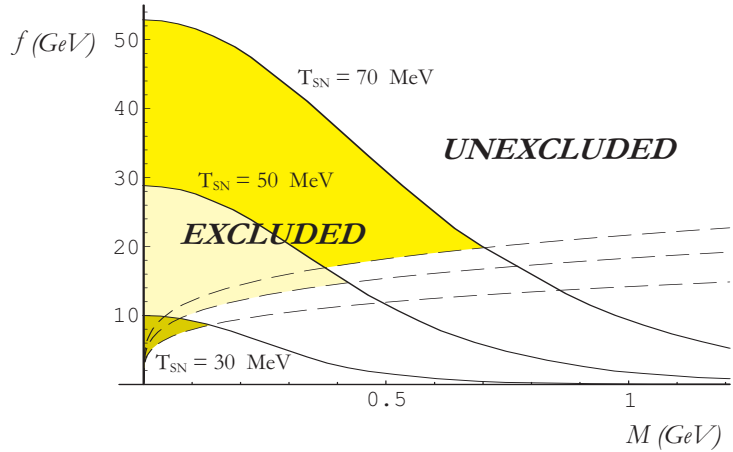,width=10.0cm}}
%\vspace*{8pt}
\caption{Contraints from supernova 1987A for a disformal scalar. The restrictions
are computed by estimating the disformal contribution to the cooling assuming
different supernova temperatures: $T_{SN}=10,\, 50,\, 70$ MeV. The solid lines
show the limits on the volume emissivity, whereas the dashed lines correspond
to the $L=10$ Km limits on the disformal mean free path \cite{CDM2}.\label{SN}}
\end{figure}

The integral in the angular variables can be performed analytically,
whereas the integral over the two energies is done numerically.
The final restrictions depend on the supernova temperature ($T_{SN}$)
and the number of disformal species ($N$). We have shown the limits
on $f$ and $M$ for  $T_{SN}=30,\,50,\,70$ MeV and for $N=1$ in Fig. \ref{SN}
\cite{CDM2}. For disformal masses of the order of the GeV, the constraints on the coupling
disappear even for $T_{SN}=70$ MeV, due to the value of the mean free path.

\section{Non-thermal disformal dark matter}
\label{nonthermals}

In the previous sections we have considered disformal
scalars as thermal DM candidates, i.e. their
primordial abundances were generated by the thermal decoupling process
in an expanding universe. However, if the reheating temperature $T_{RH}$ after inflation was sufficiently low
then the disformal fields were never in thermal
equilibrium with the plasma. However, still there is
the possibility for them to be produced non-thermally, very much in the same ways as axions \cite{axions}
or other bosonic degrees of freedom \cite{Cembranos:2012kk}. Indeed,
 if the disformal fields are understood as the pNGB associated to a global symmetry breaking from a
 group $G$ to a subgroup $H$, then they will
 correspond to the coordinates of the coset space $K=G/H$.
If $K$ is some compact space, we can denote its typical
size as $v$. In the axion case, $v$ would
correspond to the Peccei-Quinn scale $f_{PQ}$. On the
other hand, in the case of branons, it is possible to show \cite{nonthermal} that $v=f^2 R_B$ with $R_B$ the radius of the
compact extra space $B$.
Thus, if the reheating temperature was smaller than the
freeze-out temperature of the disformal fields, i.e.
$T_{RH}\ll T_f$, but larger compared to the explicit symmetry breaking scale $T_{RH}\gg (Mv)^{1/2}$, then the
disformal fields were essentially massless and
decoupled from the rest of matter fields.
In this case, we do not expect that after symmetry breaking, the initial value of the disformal field
would correspond to the minimum of the potential $\pi_0=0$,
but in general we would have $\pi_0\sim v$ within a region
of
size $H^{-1}$. The evolution of the $\pi$ fields
after symmetry breaking would then correspond to
that of a scalar field in an expanding universe, i.e., while $H\gg M$, the field is frozen in
its initial value $\pi=\pi_0$. When the temperature
falls below $T_i$ for which $3H(T_i)\simeq M$,
$\pi$ starts oscillating around the minimum with
a decaying amplitude. These oscillations correspond to a zero-momenum disformal condensate, whose energy
density scales precisely as that of CDM.

In principle, it would be possible that the disformal condensate could
transfer part of its energy to SM fields in $2\rightarrow 2$ process as those
discussed in previous sections. Asuming that $M\ll 1$ MeV and
that neutrinos are massless, it was shown in Ref.  \citen{nonthermal} that
the condition to avoid the energy depletion is $H(T_{RH})\gsim M$,
which translates into $T_{RH}\gsim (MM_P)^{1/2}$ using the Friedmann equation in a radiation dominated universe. This condition automatically ensures $T_{RH}\gg (Mv)^{1/2}$.

\begin{figure*}[ht]
\centerline{\includegraphics[trim=0cm 8cm 2cm 8cm, clip,width=10cm]{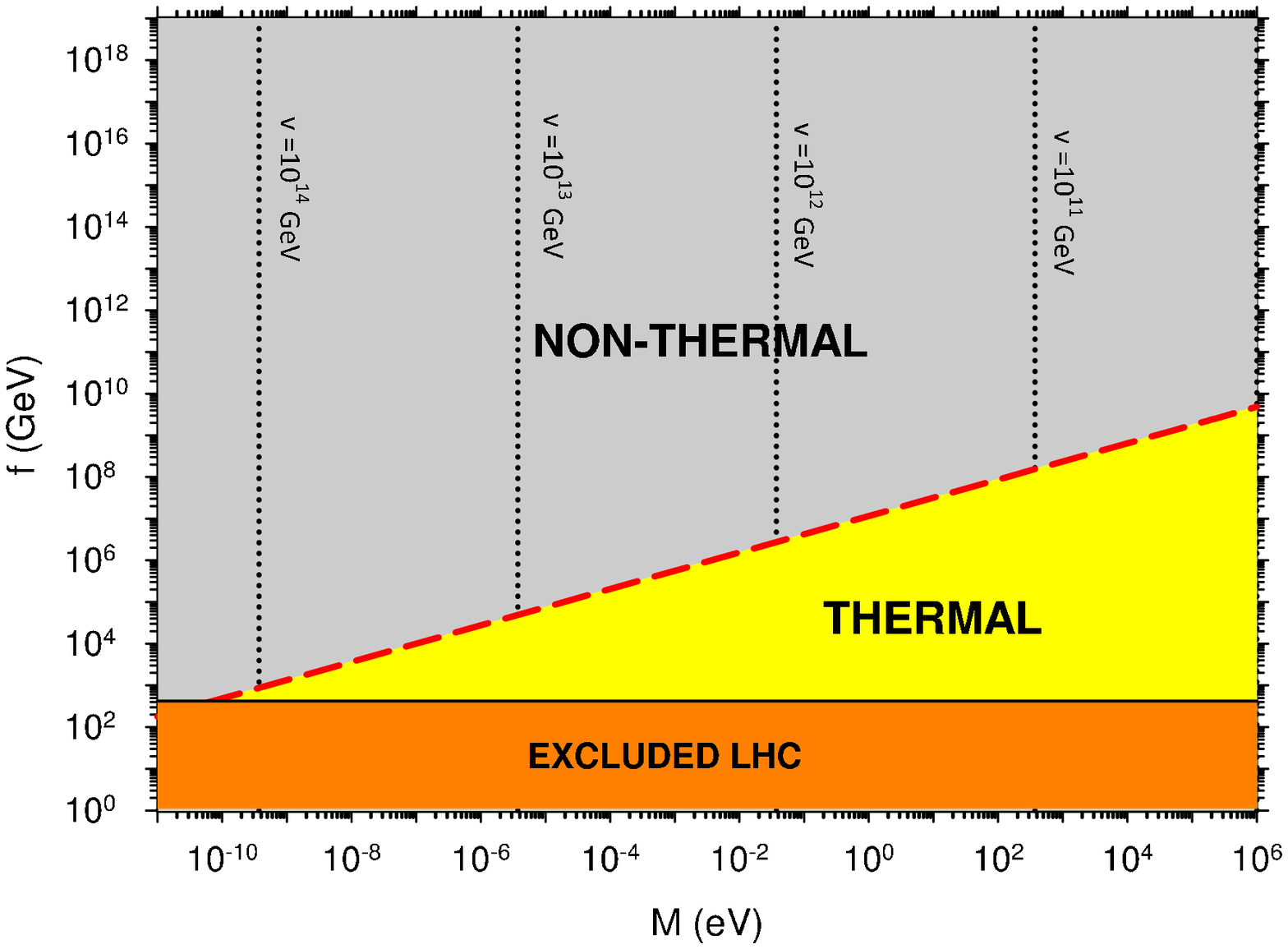}}
\caption{Thermal vs. non-thermal disformal regions in the $f - M$ plane. The dashed (red) line separates the two regions
and corresponds to $T_i=T_f$ . The dotted (black) lines
correspond to
$\Omega_{D}h^2 = 0.126 - 0.114$ \cite{Ade:2015xua}
for different values of the $v$ scale.
The regions on the right of each dotted line would be excluded by
non-thermal disformal DM overproduction. }
\label{nonthermal}
\end{figure*}

Thus we see that in order for the condensate to form and survive until present, the reheating temperature should satisfy the condition $T_i\simeq (MM_P)^{1/2}<T_{RH}< T_f$.  Therefore, if $T_i>T_f$ the interval disappears and only thermal relics are possible. In the opposite case $T_i<T_f$,
we can also have non-thermal production. As we show in Section 5, for light
disformal fields a good approximation for the freeze-out temperature
is $\log(T_f/\text{GeV})\simeq (8/7)\log(f/\text{GeV})-3.2$. In Fig. \ref{nonthermal} the $T_i=T_f$ line separating the thermal and non-thermal
regimes is plotted in the $(f,M)$ plane.

It is possible to obtain the present energy density of
the disformal field oscillations following the same
steps as in the axionic case. Assuming that $M$ does not depend on the
temperature, we find:
\begin{eqnarray}
\Omega_D h^2\simeq \frac{5 v^2 M T_0^3}{2M_P T_i\rho_c}\;,
\end{eqnarray}
with $T_0=2.75$ K the CMB temperature today and $\rho_c$ the
critical density. We can see in Fig. \ref{nonthermal}, that
for certain values of the $v$ scale, the above energy density can
be cosmologically relevant and in particular it could agree
with the measured value of the CDM abundance $\Omega_{D}h^2 = 0.126 - 0.114$  \cite{Ade:2015xua}.

\section{Conclusions}
\label{Con}

In this work, we have summarized the main phenomenology associated with disformal scalar particles.
Such a phenomenology is described by an effective action characterized by a dimension 8 interaction term with the SM
fields.
%The high dimension of this operator is a very distintive feature of these scalars.
In order to keep the stability of the high dimensionality of the leading interaction against radiative corrections
it is necessary to introduce a distinctive pattern of symmetries. These disformal symmetries lead naturally to the
stability of the disformal scalars. We have started by showing these features within a particular model associated with
flexible brane worlds. In these scenarios, the brane tension scale $f$ is much smaller than the fundamental gravitational
scale in $D$ dimensions $M_D$. Within this framework, the relevant new low-energy phenomenology is associated with
new disformal scalars called branons, which are associated with the brane oscillations along the additional spatial dimensions.

\begin{figure}[h]
%{\epsfxsize=12.0 cm \epsfbox{LogKeysoloplot.eps}}
\centerline{\psfig{file=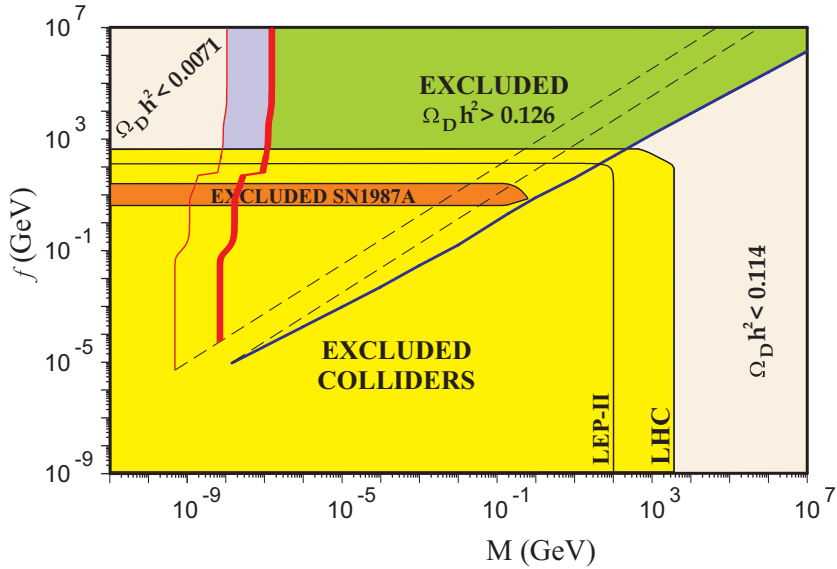,width=11.0cm}}
\vspace*{8pt}
\caption{Combined exclusion plot for a model with a single disformal scalar from total and hot DM,
LEP-II \cite{Alcaraz:2002iu,L3} and
LHC \cite{Cembranos:2004jp,LHCDirect,Khachatryan:2014rwa,Landsberg:2015pka}
single photon events,
and supernovae cooling \cite{CDM2}.
The (blue) solid line on the right is associated with cold DM behaviour.
The two (red) solid lines on the left are associated with hot DM: the
thicker one corresponds to the total DM range $\Omega_{D}h^2=0.126 - 0.114$,
and the thin one is the hot DM limit
$\Omega_{D}h^2<0.0071$ \cite{Ade:2015xua}.  The
dashed lines correspond to $x_f = 3$ for hot (upper curve) and cold (lower curve)
DM.
This figure is an updated version of the one obtained in Refs.  \citen{CDM,CDM2}.
\label{Combined}}
\end{figure}

In this model, it is straightforward to deduce the corresponding effective action and the Feynman rules relative to the
couplings of branons with SM particles. They allow to compute the cross-sections and decay rates for different processes
relevant for disformal particles production and annihilation.
In particular, production rates are necessary to establish present constraints from different particle accelerators through the analysis of missing energy and transverse momentum.
For example, for electron-positron colliders, the most sensitive channel is the single-photon one. We have used the information
coming from LEP in order to get different exclusion plots on the disformal mass $M$, and coupling $f$.
We have also completed the study for future electron-positron colliders. In any case, the current most constraining limits come
from hadron colliders. We have taken into account data obtained by HERA and  Tevatron, but recent observations by the LHC are the
most sensitive, in particular, the monophoton analysis by CMS.

There are complementary bounds on $f$ and $M$ coming from cosmology or astrophysics. It is interesting to notice that
the allowed range of parameters includes weak disformal coupling and large disfomal masses. Taking into account that disformal symmetries ensure the stability of the disformal scalars, they become natural DM candidates. Through an explicit calculation, we have shown that
the relic abundance relative to disformal particles can be cosmologically relevant and could account for the observed fraction of the abundance
in form of CDM \cite{CDM,CDM2}. From the commented effective low-energy action for massive disformal scalars, it is possible to compute the annihilation cross-sections of disformal pairs into SM particles. By solving the Boltzmann equation in an expanding universe, we can analyze the
disformal freeze-out and calculate the corresponding thermal relic abundances both for the hot and cold cases.

\begin{figure}[bt]
\begin{center}
%\resizebox{8.8cm}{6.4cm}
\resizebox{10cm}{!}
{\includegraphics{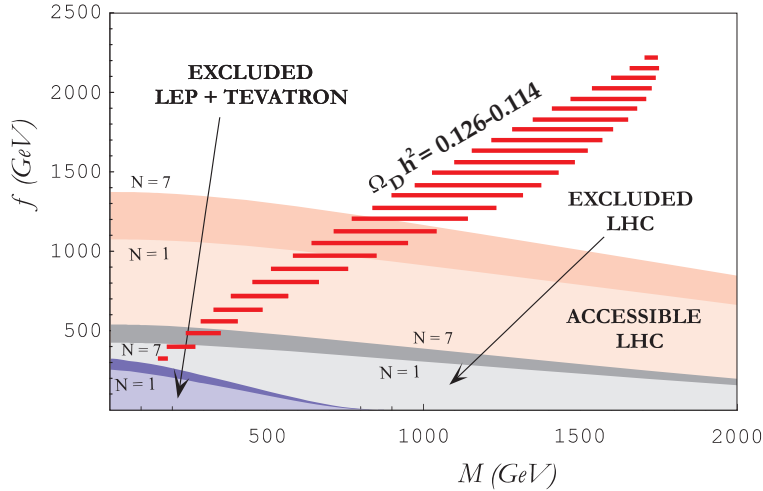}}
\caption {The shaded area shows the parameter space of disformal DM with thermal relic in the range: $\Omega_{Br}h^2=0.126 - 0.114$,
and with a contribution to the muon anomalous magnetic moment: $\delta a_\mu=(26\pm 16)\times 10^{-10}$. The lower area is
excluded by single-photon processes at LEP together with monojet signals at Tevatron \cite{BW2,Cembranos:2004jp,LHCDirect}
and from the monophoton analysis at the LHC \cite{Khachatryan:2014rwa,Landsberg:2015pka} (intermediate area).
Prospects for the sensitivity at the LHC for real branon production are plotted also for the monojet analysis for a total integral
luminosity of ${\cal L}=10^5$ and total energy in the center of mass of the collision of $\sqrt{s}=14$ TeV. The explicit dependence on
the number of disformal fields $N$ is presented, since all these regions are plotted for the extreme values $N=1$ and $N=7$.\label{forlilian1}}
\end{center}
\end{figure}

Comparing the results with the recent observational limits on the total and
hot DM energy densities, we have obtained exclusion plots in the
$f-M$ plane. Such plots are compared with the limits coming from collider
experiments. We conclude that there are essentially two allowed regions
in Fig. \ref{Combined}: one with
low disformal masses and weak disformal couplings corresponding
to hot relics, and a second region with large masses and not so weak couplings,
in which disformal particles behave as cold relics. In any case, the effective theory, which
describes the phenomenology of these disformal scalars are not valid for strongly coupled
fields. Indeed, disformal quantum effects can be parameterized by a cutoff $\Lambda$, that
limits the energy range of the model and establishes the importance of disformal loop effects
on SM phenomenology. It is interesting to note that disformal particles are able to improve the
agreement of the measured muon anomalous magnetic moment with the SM prediction (see Fig. \ref{forlilian1}).

In addition, there is an intermediate region where $f$ is comparable to $M$, which is precisely
the region studied in Refs. \citen{CDM} and \citen{CDM2}, and where disformal particles
could account for the measured cosmological DM.
This study can be also used to exclude different
regions of the parameter space of the model, where the disformal scalars are
overproduced or where they behave as hot DM avoiding a successful structure formation.

On the other hand, by using nucleosynthesis restrictions on the number of relativistic species,
we can impose an upper bound on the number of light disformal fields in terms of $f$.
If they decouple after the QCD phase transition ($f< 60$ GeV), the constraints are important
($N\leq 3$), but they become very weak otherwise. We have also discussed the possibility that
disformal scalars can contribute to the cooling of stellar objects. However, these restrictions
are not competitive with those coming from colliders.

Finally, apart from the thermal production, disformal fields can also be produced non-thermally by a similar mechanism to the axion misalignment. This fact allows to  extend to lower masses the parameter space in which this kind of fields can play the role of DM.

\section*{Acknowledgments}

This work has been supported by the Spanish MICINNs Consolider-Ingenio 2010 Programme under grant MultiDark CSD2009-00064  and MINECO grant FPA2014-53375-C2-1-P and FIS2014-52837-P.

\newpage
\appendix

\section{Vertices for disformal particles}
\label{Ver}

In this section, one can find the leading Feynman rules with outgoing momenta for massive disformal scalar particles.
They include all the interaction vertices between two disformal scalars and the SM particles with
the dependence on the momenta of the particles \cite{Alcaraz:2002iu}.

\subsection{$V1[p_1, p_2, p_3, p_4]$}

\begin{table}[h]
\begin{tabular}{c c c}
\begin{tabular}{c c c}
$\pi^\alpha(p_3)$& &$\pi^\beta(p_4)$\\
 &\epsfysize=3.0 cm\epsfxsize=3.0
cm\epsfbox{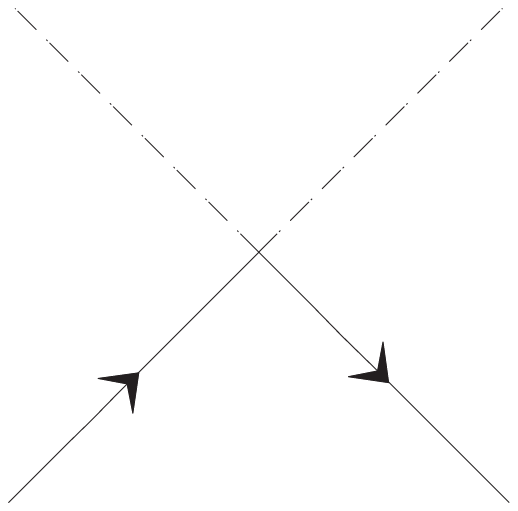}& \\
 $\bar\psi(p_1)$& &$\psi(p_2)$
\end{tabular}
 &
$\equiv$ & $V1[p_1, p_2, p_3, p_4]$
\end{tabular}
\end{table}
\begin{eqnarray}
V1&=& \frac{-i\delta^{\alpha\beta}}{4 f^4} \{\gamma^\mu
p_{4\mu}(p_3,p_1-p_2) +
\gamma^\mu p_{3\mu}(p_4,p_1-p_2)\nonumber\\
&-&\gamma^\mu
(p_{1\mu}-p_{2\mu})(\frac{3}{2}M^2+2(p_3,p_4))\nonumber\\
&+&4m_\psi((p_3,p_4)+M^2)\}.
\end{eqnarray}

\newpage
\subsection{$V2_{\mu\nu}^{ab}[p_1, p_2, p_3,p_4]$}

\begin{table}[h]
\begin{tabular}{c c c}
\begin{tabular}{c c c}
$\pi^\alpha(p_3)$& &$\pi^\beta(p_4)$\\
 &\epsfysize=3.0 cm\epsfxsize=3.0
cm\epsfbox{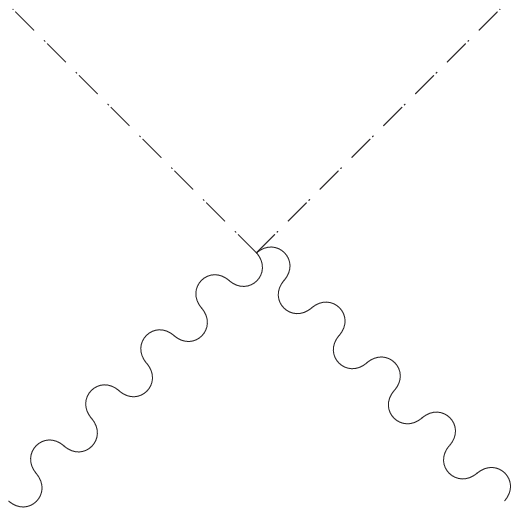}& \\
 $A^b_\nu(p_2)$& &$A^a_\mu(p_1)$
\end{tabular}
 &
$\equiv$ & $V2_{\mu\nu}^{ab}[p_1, p_2, p_3,p_4]$
\end{tabular}
\end{table}

\begin{eqnarray}
V2_{\mu\nu}^{ab}&=&
\frac{i\delta^{ab}\delta^{\alpha\beta}}{f^4}\{p_{1\nu}p_{3\mu}(p_2,p_4)+p_{3\nu}p_{2\mu}(p_1,p_4)
\nonumber\\
&+&p_{1\nu}p_{4\mu}(p_2,p_3)+p_{4\nu}p_{2\mu}(p_1,p_3)\nonumber\\
&-&\eta_{\mu\nu}((p_1,p_4)(p_2,p_3)+(p_1,p_3)(p_2,p_4)-(p_1,p_2)(p_3,p_4))\nonumber\\
&-&(p_1,p_2)(p_{4\nu}p_{3\mu}+p_{3\nu}p_{4\mu})-(p_3,p_4)(p_{1\nu}p_{2\mu})\nonumber\\
&-&\frac{1}{2}M_a^2(2p_{4\nu}p_{3\mu}+2p_{3\nu}p_{4\mu}
-2\eta_{\mu\nu}(p_3,p_4)-\eta_{\mu\nu}M^2)\}.\nonumber\\
\end{eqnarray}
Here, the flat background metric has been used in order
to contract indices within the Fadeev-Popov action.

\newpage
\subsection{$V3^a_\mu[p_3,p_4]$}
\begin{table}[h]
\begin{tabular}{c c c}
\begin{tabular}{c c c}
\hspace*{0cm}\vspace*{2cm}$\pi^\alpha(p_3)$& &\hspace*{-2cm}$\pi^\beta(p_4)$\\
%\vspace{length}
\hspace*{-2cm}\vspace*{-2.5cm}$\bar\psi(p_1)$& &$\psi(p_2)$\\
&\hspace*{-1.2cm}\epsfysize=3.0
cm\epsfxsize=3.0cm\epsfbox{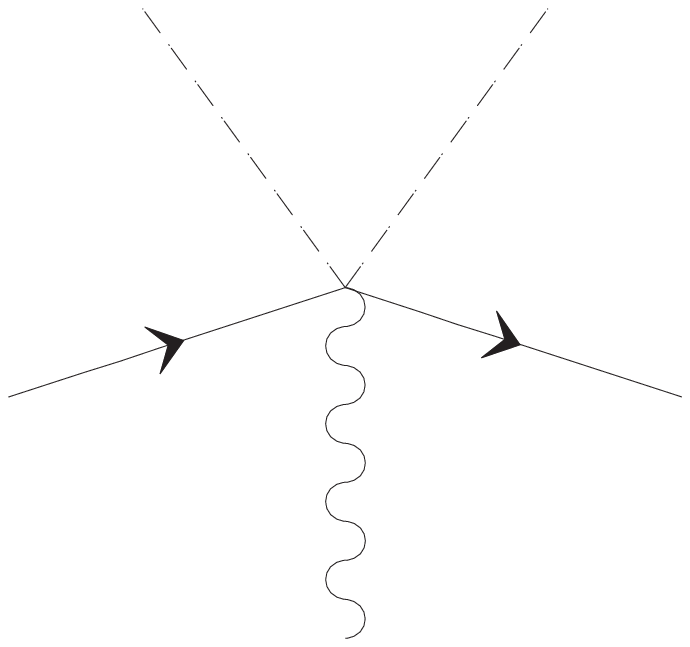}&
 \\
&\hspace*{-1.2cm}$A^a_\mu(p_5)$&
\end{tabular}
 &
$\equiv$ & $V3^a_\mu[p_3,p_4]$
\end{tabular}
\end{table}

\begin{eqnarray}
V3^a_\mu&=& \frac{-hT^a\delta^{\alpha\beta}}{4 f^4} \{2\gamma^\nu
p_{4\nu}p_{3\mu} +
2\gamma^\nu p_{3\nu}p_{4\mu}\nonumber\\
&+&\gamma_\mu(-3M^2 - 4 (p_3, p_4)\}(c_V-c_A\gamma_5).
\end{eqnarray}

\subsection{$V4_{\mu\nu\lambda}^{abc}[p_1,p_2,p_3,p_4,p_5]$}

\begin{table}[h]
\begin{tabular}{c c c}
\begin{tabular}{c c c}
\hspace*{0cm}\vspace*{2cm}$\pi^\alpha(p_3)$& &\hspace*{-2cm}$\pi^\beta(p_4)$\\
%\vspace{length}
\hspace*{-2cm}\vspace*{-2.5cm}$A^b_\mu(p_1)$& &$A^c_\nu(p_2)$\\
&\hspace*{-1.2cm}\epsfysize=3.0 cm\epsfxsize=3.0
cm\epsfbox{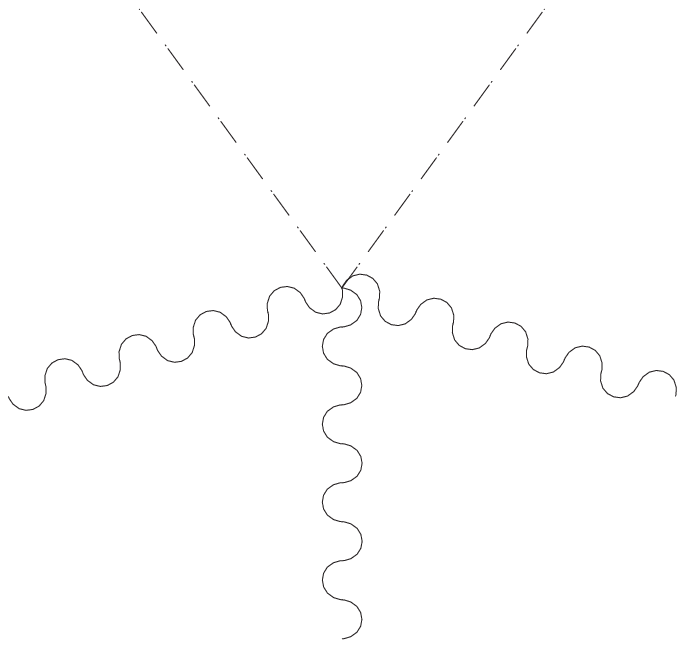}&
 \\
&\hspace*{-1.2cm}$A^a_\lambda(p_5)$&
\end{tabular}
 &
$\equiv$ & $V4_{\mu\nu\lambda}^{abc}[p_1,p_2,p_3,p_4,p_5]$
\end{tabular}
\end{table}

\begin{eqnarray}
 V4_{\mu\nu\lambda}^{abc}&=&\frac{h
C^{abc}\delta^{\alpha\beta}}{f^4}\{p_{1\nu}(p_{3\mu}p_{4\lambda}+p_{3\lambda}p_{4\mu})-p_{1\lambda}(p_{3\mu}p_{4\nu}+p_{3\nu}p_{4\mu})\\
&+&p_{2\lambda}(p_{3\nu}p_{4\mu}+p_{3\mu}p_{4\nu})-p_{2\mu}(p_{3\nu}p_{4\lambda}+p_{3\lambda}p_{4\nu})\nonumber\\
&+&p_{5\mu}(p_{3\nu}p_{4\lambda}+p_{3\lambda}p_{4\nu})-p_{5\nu}(p_{3\mu}p_{4\lambda}+p_{3\lambda}p_{4\mu})\nonumber\\
&+&\eta_{\lambda\nu}((p_3,p_4)(p_{2\mu}-p_{5\mu})+p_{4\mu}(p_5-p_2,p_3)+p_{3\mu}(p_5-p_2,p_4))\nonumber\\
&+&\eta_{\lambda\mu}((p_3,p_4)(p_{5\nu}-p_{1\nu})+p_{4\nu}(p_1-p_5,p_3)+p_{3\nu}(p_1-p_5,p_4))\nonumber\\
&+&\eta_{\mu\nu}((p_3,p_4)(p_{1\lambda}-p_{2\lambda})+p_{4\lambda}(p_2-p_1,p_3)+p_{3\lambda}(p_2-p_1,p_3))\}\;.\nonumber
\end{eqnarray}

\newpage

\subsection{$V5[p_1, p_2, p_3, p_4]$}

\begin{table}[h]
\begin{tabular}{c c c}
\begin{tabular}{c c c}
$\pi^\alpha(p_3)$& &$\pi^\beta(p_4)$\\
 &\epsfysize=3.0 cm\epsfxsize=3.0
cm\epsfbox{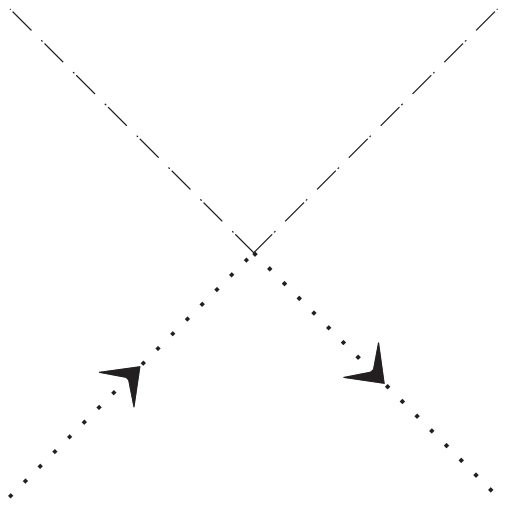}& \\
 $\phi^\dagger(p_1)$& &$\phi(p_2)$
\end{tabular}
 &
$\equiv$ & $V5[p_1, p_2, p_3, p_4]$
\end{tabular}
\end{table}
\begin{eqnarray}
V5&=& \frac{i\delta^{\alpha\beta}}{f^4} \left\{-[(p_3,p_4)+M^2]
[(p_1,p_2)+m_\phi^2]\right.\nonumber\\
&+&\left.(p_1,p_3)(p_4,p_2)+(p_2,p_3)(p_4,p_1)
+\frac{1}{2}M^2(p_1,p_2)\right\}.
\end{eqnarray}

\newpage
\subsection{$V6^a_\mu[p_3,p_4]$}
\begin{table}[h]
\begin{tabular}{c c c}
\begin{tabular}{c c c}
\hspace*{0cm}\vspace*{2cm}$\pi^\alpha(p_3)$& &\hspace*{-2cm}$\pi^\beta(p_4)$\\
%\vspace{length}
\hspace*{-2cm}\vspace*{-2.5cm}$\phi^\dagger(p_1)$& &$\phi(p_2)$\\
&\hspace*{-1.2cm}\epsfysize=3.0
cm\epsfxsize=3.0cm\epsfbox{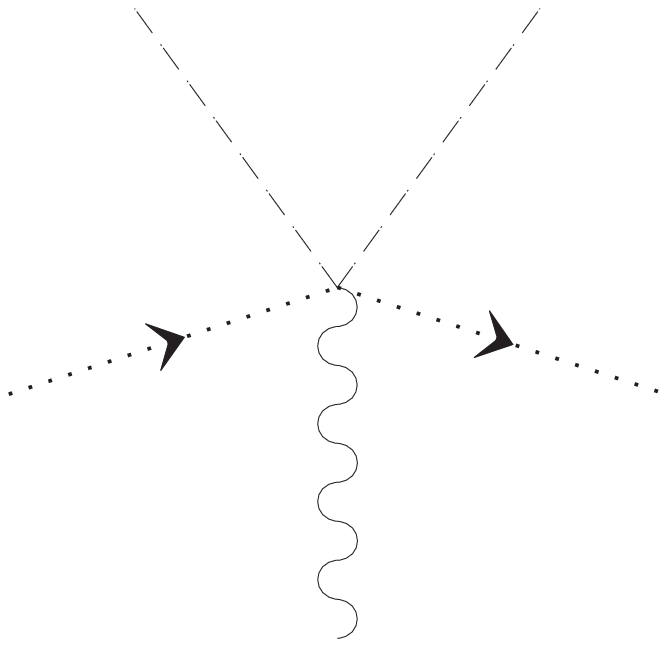}&
 \\
&\hspace*{-1.2cm}$A^a_\mu(p_5)$&
\end{tabular}
 &
$\equiv$ & $V6^a_\mu[p_3,p_4]$
\end{tabular}
\end{table}

\begin{eqnarray}
V6^a_\mu&=& \frac{hT^a\delta^{\alpha\beta}}{f^4} \{(p_{1}-p_{2})_\mu[(p_3,
p_4) +
\frac{1}{2}M^2]\nonumber\\
&+&(p_1-p_2,p_3)p_{4\mu}+(p_1-p_2,p_4)p_{3\mu}\}.
\end{eqnarray}

\subsection{$V7^{ab}_{\mu\nu}[p_3,p_4]$}
\begin{table}[h]
\begin{tabular}{c c c}
\begin{tabular}{c c c}
\hspace*{0cm}\vspace*{1.2cm}$\pi^\alpha(p_3)$& &\hspace*{-2cm}$\pi^\beta(p_4)$\\
%\vspace{length}
\hspace*{-2cm}\vspace*{-1.7cm}$\phi^\dagger(p_1)$& &$\phi(p_2)$\\
&\hspace*{-1.2cm}\epsfysize=3.0
cm\epsfxsize=3.0cm\epsfbox{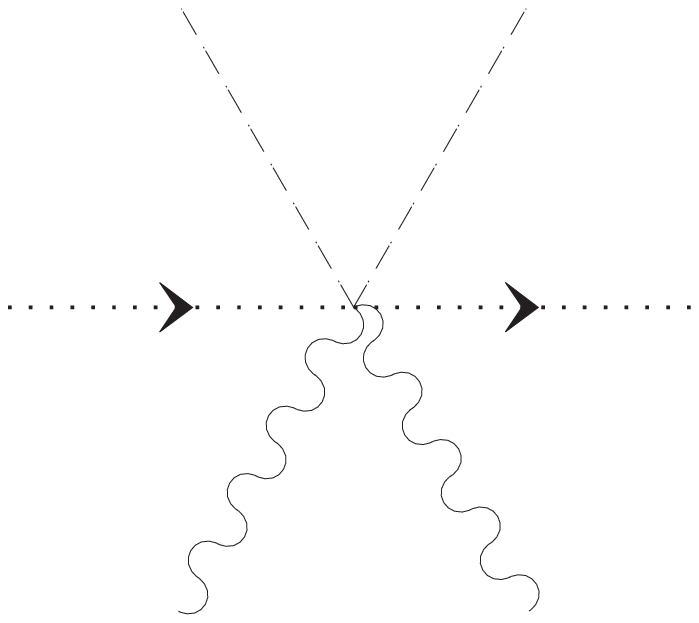}&
 \\
\hspace*{0cm}$A^a_\mu(p_5)$& &\hspace*{-2cm}$A^b_\nu(p_6)$
\end{tabular}
 &
$\equiv$ & $V7^{ab}_{\mu\nu}[p_3,p_4]$
\end{tabular}
\end{table}

\begin{eqnarray}
V7^{ab}_{\mu\nu}&=& \frac{-ih^2\{T^a,T^b\}\delta^{\alpha\beta}}{f^4} \{[(p_3, p_4)
+ \frac{1}{2}M^2]\eta_{\mu\nu}
%&+&
-p_{3\mu} p_{4\nu}-p_{4\mu} p_{3\nu}\}.
\end{eqnarray}
\newpage
\subsection{$V8^{abcd}_{\mu\nu\rho\sigma}[p_3,p_4]$}
\begin{table}[h]
\begin{tabular}{c c c}
\begin{tabular}{c c c}
\hspace*{0cm}\vspace*{1.2cm}$\pi^\alpha(p_3)$& &\hspace*{-2cm}$\pi^\beta(p_4)$\\
%\vspace{length}
\hspace*{-2cm}\vspace*{-1.7cm}$A^c_\rho(p_5)$& &$A^d_\sigma(p_6)$\\
&\hspace*{-1.2cm}\epsfysize=3.0
cm\epsfxsize=3.0cm\epsfbox{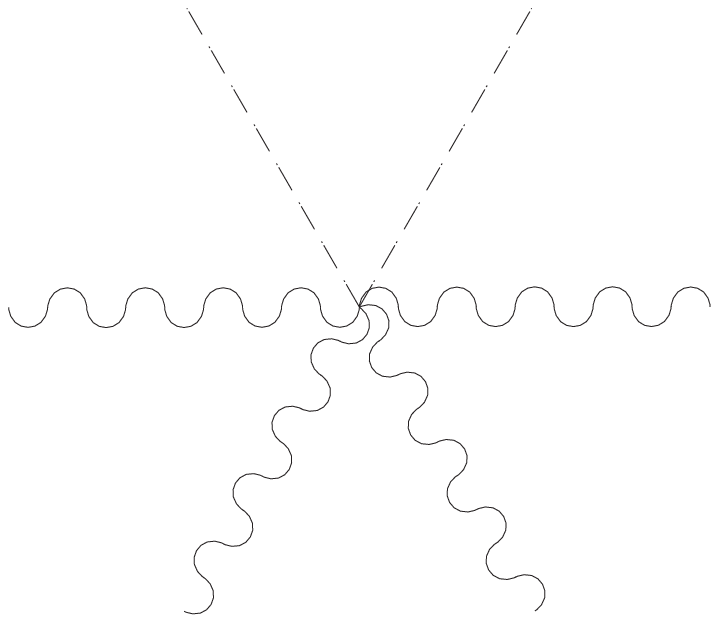}&
 \\
\hspace*{0cm}$A^a_\mu(p_1)$& &\hspace*{-2cm}$A^b_\nu(p_2)$
\end{tabular}
 &
$\equiv$ & $V8^{abcd}_{\mu\nu\rho\sigma}[p_3,p_4]$
\end{tabular}
\end{table}

\begin{eqnarray}
V8^{abcd}_{\mu\nu\rho\sigma}[p_3,p_4]&=&
\frac{ih^2\delta^{\alpha\beta}}{f^4}
\{C^{eab}C^{ecd}[\eta_{\nu\sigma}(p_{3\rho}p_{4\mu}+p_{3\mu}p_{4\rho})\nonumber\\
&-&
\eta_{\nu\rho}(p_{3\sigma}p_{4\mu}+p_{3\mu}p_{4\sigma})\nonumber\\
&+&
\eta_{\mu\sigma}(\eta_{\nu\rho}(p_3,p_4)-p_{3\rho}p_{4\nu}-p_{3\nu}p_{4\rho})\nonumber\\
&-&
\eta_{\mu\rho}(\eta_{\nu\sigma}(p_3,p_4)-p_{3\sigma}p_{4\nu}-p_{3\nu}p_{4\sigma})]\nonumber\\
&+&
C^{eac}C^{ebd}[\eta_{\nu\mu}(p_{3\rho}p_{4\sigma}+p_{3\sigma}p_{4\rho})\nonumber\\
&-&
\eta_{\nu\rho}(p_{3\sigma}p_{4\mu}+p_{3\mu}p_{4\sigma})\nonumber\\
&+&
\eta_{\mu\sigma}(\eta_{\nu\rho}(p_3,p_4)-p_{3\rho}p_{4\nu}-p_{3\nu}p_{4\rho})\nonumber\\
&-&
\eta_{\sigma\rho}(\eta_{\nu\mu}(p_3,p_4)-p_{3\mu}p_{4\nu}-p_{3\nu}p_{4\mu})]\nonumber\\
&+&
C^{ead}C^{ebc}[\eta_{\nu\mu}(p_{3\rho}p_{4\sigma}+p_{3\sigma}p_{4\rho})\nonumber\\
&-&
\eta_{\nu\sigma}(p_{3\rho}p_{4\mu}+p_{3\mu}p_{4\rho})\nonumber\\
&+&
\eta_{\mu\rho}(\eta_{\nu\sigma}(p_3,p_4)-p_{3\sigma}p_{4\nu}-p_{3\nu}p_{4\sigma})\nonumber\\
&-&
\eta_{\sigma\rho}(\eta_{\nu\mu}(p_3,p_4)-p_{3\mu}p_{4\nu}-p_{3\nu}p_{4\mu})]\}.
\end{eqnarray}

\newpage

\section{Production and
annihilation cross-sections of disformal scalars
with SM particles}
\label{Cross}

In this appendix, the production and annihilation cross-sections of disformal particles in processes involving SM particles
are presented. They were computed in Refs. \citen{CDM,CDM2} by using the Feynman rules detailed in the previous section \cite{Alcaraz:2002iu}.
$N$ is the number of disformal scalars. The internal degrees of freedom are summed for final particles, whereas they are averaged for initial ones.

\subsection{Scalars}

\centerline{$\sigma_1$:
$\Phi^\dagger(p_1),\Phi(p_2)\longrightarrow\pi(p_3),\pi(p_4)$}
\begin{eqnarray}
\sigma_1&=&\frac{N}{7680
f^8\pi s}\sqrt{\frac{(s-4M^2)}{(s-4m_\Phi^2)}}[-s(8m_\Phi^2+s)(s-4M^2)^2\nonumber\\
&&+(2m_\Phi^2+s)^2(23M^4-14M^2s+3s^2)].
\end{eqnarray}

\vspace{.5cm}

\centerline{$\sigma_2$: $\pi(p_1),\pi(p_2)\longrightarrow
\Phi^\dagger(p_3),\Phi(p_4)$}
\begin{eqnarray}
\sigma_2&=&\frac{1}{3840 N f^8\pi s}\sqrt{\frac{(s-4m_\Phi^2)}
{(s-4M^2)}}[-s(8m_\Phi^2+s)(s-4M^2)^2\nonumber\\
&&+(2m_\Phi^2+s)^2(23M^4-14M^2s+3s^2)].
\end{eqnarray}

These cross-sections correspond to a complex scalar such as charged mesons.
For a real degree of freedom, like a neutral meson or the Higgs field,
the annihilation cross-section of disformal particles has to be divided by two.

\subsection{Fermions}

\centerline{$\sigma_3$:
$\psi^+(p_1),\psi^-(p_2)\longrightarrow\pi(p_3),\pi(p_4)$}
\begin{eqnarray}
\sigma_3&=&\frac{N}{30720
f^8\pi}\sqrt{(s-4M^2)(s-4m_{\psi}^2)}[(s-4M^2)^2\nonumber\\
&&+\frac{2m_{\psi}^2}{s}(23M^4-14M^2s+3s^2)].
\end{eqnarray}

\vspace{.5cm}

\centerline{$\sigma_4$: $\pi(p_1),\pi(p_2)\longrightarrow
\psi^+(p_3),\psi^-(p_4)$}

\begin{eqnarray}
\sigma_4&=&\frac{1}{3840
N f^8\pi}\sqrt{\frac{(s-4m_{\psi}^2)^3}{(s-4M^2)}}[(s-4M^2)^2\nonumber \\
&+&\frac{2m_{\psi}^2}{s}(23M^4-14M^2s+3s^2)].
\end{eqnarray}

These cross-sections are associated with Dirac fermion of mass $m_{\psi}$.
For the case of a massless Weyl fermion, the production cross-section of disformal pairs has to be
multiply by two, whereas the annihilation cross-section has to be divided by two.

\subsection{Photons}

\centerline{$\sigma_5$:
$\gamma(p_1),\gamma(p_2)\longrightarrow\pi(p_3),\pi(p_4)$}
\begin{eqnarray}
\sigma_5&=&\frac{N}{7680
f^8\pi}\sqrt{1-\frac{4M^2}{s}}s(s-4M^2)^2.
\end{eqnarray}

\vspace{.5cm}

\centerline{$\sigma_6$:
$\pi(p_1),\pi(p_2)\longrightarrow\gamma(p_3),\gamma(p_4)$}
\begin{eqnarray}
\sigma_6&=&\frac{1}{1920 N
f^8\pi}\frac{s(s-4M^2)^2}{\sqrt{1-\frac{4M^2}{s}}}.
\end{eqnarray}

\subsection{$ZZ$}
\centerline{$\sigma_7$:
$Z(p_1),Z(p_2)\longrightarrow\pi(p_3),\pi(p_4)$}
\begin{eqnarray}
\sigma_7&=&\frac{N}{69120 f^8\pi s} \sqrt{\frac{s-4M^2}{s-4M_Z^2}}
[3s(8M_Z^2+s)(s-4M^2)^2\nonumber\\
&&+(12M_Z^4+4sM_Z^2+s^2)(23M^4-14M^2s+3s^2)].
\end{eqnarray}

\vspace{.5cm}

\centerline{$\sigma_8$: $\pi(p_1),\pi(p_2)\longrightarrow
Z(p_3),Z(p_4)$}
\begin{eqnarray}
\sigma_8&=&\frac{1}{7680 N f^8\pi s}
\sqrt{\frac{s-4M_Z^2}{s-4M^2}}
[3s(8M_Z^2+s)(s-4M^2)^2\nonumber\\
&&+(12M_Z^4+4sM_Z^2+s^2)(23M^4-14M^2s+3s^2)].
\end{eqnarray}

\subsection{$W^+W^-$}
\centerline{$\sigma_9$:
$W^\pm(p_1),W^\mp(p_2)\longrightarrow\pi(p_3),\pi(p_4)$}
\begin{eqnarray}
\sigma_9&=&\frac{N}{69120 f^8\pi s} \sqrt{\frac{s-4M^2}{s-4M_W^2}}
[3s(8M_W^2+s)(s-4M^2)^2\nonumber\\
&&+(12M_W^4+4sM_W^2+s^2)(23M^4-14M^2s+3s^2)].
\end{eqnarray}

\vspace{.5cm}

\centerline{$\sigma_{10}$: $\pi(p_1),\pi(p_2)\longrightarrow
W^\pm(p_3),W^\mp(p_4)$}
\begin{eqnarray}
\sigma_{10}&=&\frac{2}{7680 N f^8\pi s}
\sqrt{\frac{s-4M_W^2}{s-4M^2}}
[3s(8M_W^2+s)(s-4M^2)^2\nonumber\\
&&+(12M_W^4+4sM_W^2+s^2)(23M^4-14M^2s+3s^2)].
\end{eqnarray}

\subsection{Gluons}
\centerline{$\sigma_{11}$:
$g(p_1),g(p_2)\longrightarrow\pi(p_3),\pi(p_4)$}
\begin{eqnarray}
\sigma_{11}&=&\frac{N}{61440
f^8\pi}\sqrt{1-\frac{4M^2}{s}}s(s-4M^2)^2.
\end{eqnarray}

\vspace{.5cm}

\centerline{$\sigma_{12}$: $\pi(p_1),\pi(p_2)\longrightarrow
g(p_3),g(p_4)$}
\begin{eqnarray}
\sigma_{12}&=&\frac{1}{240 N
f^8\pi}\frac{s(s-4M^2)^2}{\sqrt{1-\frac{4M^2}{s}}}.
\end{eqnarray}

\section{Thermal averages of disformal annihilation cross-sections}
\label{Thermal}

In this appendix, the thermal average of the annihilation cross-section $\langle \sigma_A v\rangle$ of disformal scalars
into SM particles are summarized for the different contributing channels. \cite{CDM,CDM2}
The corresponding complete annihiliation (and production) cross-section are detailed in the previous section.
Here, the expanded the expressions for each particle species in powers of $1/x$ are presented for cold relics:
\begin{eqnarray}
\langle \sigma_A v\rangle=c_0
+c_1\frac{1}{x}+c_2\frac{1}{x^2}+
\Od(x^{-3})\;.
\end{eqnarray}

In the case of hot disformal scalars, we  give the results for the different contributions to the annihilation
rate $\Gamma_A=n_{eq}\langle\sigma_A v \rangle$, where the ultrarrelativistic behaviour is assumed ($M=0$).
In order to illustrate the high-temperature limit, we take zero mass for SM particles.

\subsection{Dirac fermions}
\subsubsection*{$x\gg 3$ (Cold)}
%\begin{eqnarray}
%\langle \sigma_A^{Dirac} v\rangle=c_0
%+c_1\frac{1}{x}+c_2\frac{1}{x^2}+
%\Od(x^{-3})
%\end{eqnarray}
%where
\begin{eqnarray}
c_0&=&\frac{1}{16\pi^2f^8}M^2 m_\psi^2(M^2-m_\psi^2)
\sqrt{1-\frac{m_\psi^2}{M^2}}\;,\\
c_1&=&\frac{1}{192\pi^2f^8}M^2 m_\psi^2(67M^2-31m_\psi^2)
\sqrt{1-\frac{m_\psi^2}{M^2}}\;,\\
c_2&=&\frac{1}{7680\pi^2f^8}\frac{M^2}{M^2-m_\psi^2}
(17408M^6+13331M^4 m_\psi^2\nonumber \\
&-&46606M^2 m_\psi^4
+18927 m_\psi^6)
\sqrt{1-\frac{m_\psi^2}{M^2}}\;.
\end{eqnarray}
This type of expansions are not valid for disformal masses close to a SM particle mass.
We can see that annihilation mainly takes place through $s$-wave since the $c_0$ coefficient is different from zero.
\subsubsection*{$x\ll 3$ (Hot)}
In this case, for massless fermions:
\begin{eqnarray}
\Gamma_A^{Dirac}=\frac{8\pi^9 T^9}{297675 \zeta(3)f^8}+\Od(x)\;.
\end{eqnarray}
\subsection{Massive gauge field}
\subsubsection*{$x\gg 3$ (Cold)}
\begin{eqnarray}
c_0&=&
\frac{M^2\,{\sqrt{1 - \frac{{m_Z}^2}{M^2}}}\,
    \left( 4\,M^4 - 4\,M^2\,{m_Z}^2 + 3\,
{m_Z}^4 \right) }{64\,f^8\,{\pi }^2}\;,\\
c_1&=&\frac{M^2\,{\sqrt{1 - \frac{{m_Z}^2}{M^2}}}\,
    \left( 364\,M^6 - 584\,M^4\,{m_Z}^2 + 349\,M^2\,
{m_Z}^4 - 93\,{m_Z}^6 \right) }{768\,
    f^8\,\left( M^2 - {m_Z}^2 \right) \,{\pi }^2}\;,\nonumber\\
c_2&=&\frac{M^2\,{\sqrt{1 - \frac{{m_Z}^2}{M^2}}}\,
     }{30720\,f^8\,
    {\left( M^2 - {m_Z}^2 \right) }^2\,{\pi }^2}
\left( 415756\,M^8 - 755844\,M^6\,{m_Z}^2\right. \nonumber\\
&+& \left. 356541\,M^4\,
{m_Z}^4 -
      76294\,M^2\,{m_Z}^6 + 56781\,{m_Z}^8 \right)\;.\nonumber
\end{eqnarray}
Similarly to the previous case, this expression is not valid near SM particles thresholds, and the dominating term
is associated with $s$-wave annihilation.
\subsubsection*{$x\ll 3$ (Hot)}
For $T\gg m_Z$, one can obtain the following leading term:
\begin{eqnarray}
\Gamma_A^{Z}=\frac{8\pi^9
T^9}{99225 \zeta(3)f^8}+\Od(x)\;.
\end{eqnarray}
\subsection{Massless gauge field}
\subsubsection*{$x\gg 3$ (Cold)}
\begin{eqnarray}
c_0&=&0\;,\\
c_1&=&0\;,\nonumber\\
c_2&=&\frac{68\,M^6}{15\,f^8\,{\pi }^2}\;.\nonumber
\end{eqnarray}
Here, $c_0=c_1=0$, which implies that the dominant term corresponds to $d$-wave annihilation.
\subsubsection*{$x\ll 3$ (Hot)}
\begin{eqnarray}
\Gamma_A^{\gamma}=\frac{16\pi^9 T^9}{297675 \zeta(3)f^8}+\Od(x)\;.
\end{eqnarray}
\subsection{Complex scalar field}
\subsubsection*{$x\gg 3$ (Cold)}
\begin{eqnarray}
c_0&=&\frac{M^2\,{\left( 2\,M^2 + {m_\Phi}^2 \right) }^2\,
{\sqrt{1 - \frac{{m_\Phi}^2}{M^2}}}}{32\,f^8\,{\pi }^2}\;,\\
c_1&=&\frac{M^2\,\left( 2\,M^2 + {m_\Phi}^2 \right)
\,{\sqrt{1 - \frac{{m_\Phi}^2}{M^2}}}\,
    \left( 182\,M^4 - 115\,M^2\,{m_\Phi}^2
- 31\,{m_\Phi}^4 \right) }{384\,f^8\,
    \left( M^2 - {m_\Phi}^2 \right) \,{\pi }^2}\;,\nonumber\\
c_2&=&\frac{M^2\,{\sqrt{1 - \frac{{m_\Phi}^2}{M^2}}}\,
     }{5120\,f^8
\,{\left( M^2 - {m_\Phi}^2 \right) }^2\,{\pi }^2}
\left( 92164\,M^8 - 123556\,M^6\,{m_\Phi}^2 \right.\nonumber \\
&+& \left. 12269\,M^4\,{m_\Phi}^4 + 9754\,M^2\,{m_\Phi}^6 +
      6309\,{m_\Phi}^8 \right)\;.\nonumber
\end{eqnarray}
This expression is reliable for disformal masses not near SM particles thresholds. In such a case, we can see that the dominant contribution is $s$-wave.
\subsubsection*{$x\ll 3$ (Hot)}
In the case of a massless scalar:
\begin{eqnarray}
\Gamma_A^{\Phi}=\frac{16\pi^9 T^9}{297675 \zeta(3)f^8}+\Od(x)\;.
\end{eqnarray}
The above results have to be divided by two for the case of a real scalar field.

It is interesting to note  that for conformal matter
such as massless fermions or gauge fields, the leading contribution is the $d$-wave.
Indeed, the $d$-term is the genuine disformal behaviour since it is provided by the disformal coupling.
This is related to the fact that for fermions, in the massless limit, the leading contribution is also
$d$-wave, whereas it is not the case for minimally coupled massless scalar fields or taking the $m_Z\rightarrow 0$
limit for massive gauge bosons.

%\begin{thebibliography}{000} %for 3 digits
%\begin{thebibliography}{00}  %for 2 digits

\end{document}